\documentclass[twocolumn,twocolappendix]{aastex63}

\usepackage{amsmath}
\usepackage{soul}
\definecolor{dgreen}{RGB}{26,148,49}
\definecolor{forest}{RGB}{3, 148, 49}

\shorttitle{Data-driven Cosmology from 3D Light Cones}
\shortauthors{Cheng et al.}
\graphicspath{{./}{figures/}}

\begin{document}

\title{Data-driven Cosmology from Three-dimensional Light Cones}

\author[0000-0002-5437-0504]{Yun-Ting Cheng}
\address{California Institute of Technology, 1200 E. California Boulevard, Pasadena, CA 91125, USA}
\email{ycheng3@caltech.edu}

\author[0000-0002-5854-8269]{Benjamin D. Wandelt}
\address{Sorbonne Universit{\'e}, CNRS, UMR 7095, Institut d’Astrophysique de Paris, 98 bis bd Arago, 75014 Paris, France}
\address{Center for Computational Astrophysics, Flatiron Institute, 162 5th Avenue, New York, NY 10010, USA}

\author[0000-0001-5929-4187]{Tzu-Ching Chang}
\address{Jet Propulsion Laboratory, California Institute of Technology, 4800 Oak Grove Drive, Pasadena, CA 91109, USA}
\address{California Institute of Technology, 1200 E. California Boulevard, Pasadena, CA 91125, USA}

\author[0000-0001-7432-2932]{Olivier Dor{\'e}}
\address{Jet Propulsion Laboratory, California Institute of Technology, 4800 Oak Grove Drive, Pasadena, CA 91109, USA}
\address{California Institute of Technology, 1200 E. California Boulevard, Pasadena, CA 91125, USA}

\begin{abstract}
We present a data-driven technique to analyze multifrequency images from upcoming cosmological surveys mapping large sky area. Using full information from the data at the two-point level, our method can simultaneously constrain the large-scale structure (LSS), the spectra and redshift distribution of emitting sources, and the noise in the observed data without any prior assumptions beyond the homogeneity and isotropy of cosmological perturbations. In particular, the method does not rely on source detection or photometric or spectroscopic redshift estimates. Here, we present the formalism and demonstrate our technique with a mock observation from nine optical and near-infrared photometric bands. Our method can recover the input signal and noise without bias, and quantify the uncertainty on the constraints. Our technique provides a flexible framework to analyze the LSS observation traced by different types of sources, which has potential for  wide application to current or future cosmological datasets such as SPHEREx, Rubin Observatory, Euclid, or the Nancy Grace Roman Space Telescope.
\end{abstract}


\section{Introduction} \label{S:intro}

The large-scale structure (LSS) of the universe is one of the most important probes of cosmology. While generations of cosmic microwave background (CMB) fluctuations measurements have provided powerful constraints on the initial conditions of the LSS \citep{2013ApJS..208...19H, 2020A&A...641A...6P, 2020JCAP...12..047A}, the late-time LSS evolution also contains information on crucial components of the current cosmological model, such as dark matter, dark energy, and primordial non-Gaussianity. This  motivated many of the  large-scale galaxy survey programs over the past decades, including 2dFGRS \citep{2005MNRAS.362..505C}, SDSS \citep{2006PhRvD..74l3507T}, WiggleZ \citep{2012PhRvD..86j3518P}, BOSS \citep{2017MNRAS.470.2617A}, eBOSS \citep{2021PhRvD.103h3533A}, KiDS \citep{2021A&A...646A.140H}, DES \citep{2018PhRvD..98d2006E, 2022PhRvD.105b3520A}, and HSC \citep{2018PASJ...70S...4A}. A next-generation of ambitious cosmological surveys already or about to come online include DESI \citep{2016arXiv161100036D}, Euclid \citep{2011arXiv1110.3193L}, the Rubin Observatory LSST \citep{2009arXiv0912.0201L}, the Nancy Grace Roman Space Telescope \citep{2015arXiv150303757S}, PFS \citep{2016SPIE.9908E..1MT}, and SPHEREx \citep{2014arXiv1412.4872D, 2018arXiv180505489D}. 

These observations are mostly designed to probe the LSS with an individual galaxy detection approach. They map the distribution of individual resolved galaxies to trace the underlying matter density field, and then infer cosmological information from the LSS clustering. With larger angular and spectral coverage, improved sensitivity and/or spectral resolution, upcoming surveys will map LSS at unprecedented line-of-sight distances and angular scales. However, as we push the observing frontier toward higher redshift, the conventional galaxy detection approach becomes suboptimal, since only the brightest objects at high redshift can stand out from noise and foregrounds and thus be detected individually. For example, \citet{2019ApJ...877...86C} showed that in the low signal-to-noise ratio regime, galaxy detection will not optimally trace the underlying LSS signal. In addition, as galaxy detection only probes bright sources above the detection limit, a substantial amount of information from fainter sources is lost. These considerations motivate the development of alternative analysis methods for upcoming LSS surveys to fully exploit the wealth of information they contain. We aim at capturing the information not only from the bright (detected) sources but also from the faint (unresolved) sources emitting in the diffuse ``spectral intensity maps'', i.e., intensity maps in all observed spectral bands.

Here, we present a novel analysis framework to fully exploit the LSS information in a 3D light cone. With minimal assumptions, our technique provides a framework to study the 3D LSS with all photons measured in spectral-intensity maps without resolving individual sources. In spectral-intensity maps, the 3D spatial distribution of emitting sources follows the underlying clustering of LSS on large scales, and there exists a well-known linear encoding scheme projecting the source spectral energy distribution (SED), redshift distribution, and the 3D large-scale clustering to the 2D spectral-intensity space. We explicit this relation in both 2D projected map space and in correlation space, i.e. the cross angular power spectrum $C_{\ell, \nu\nu'}$ for all combinations of observed frequencies $\nu$ and $\nu'$. For a given observed angular power spectrum $C_{\ell, \nu\nu'}$, we formalize the likelihood function on the underlying source SED, redshift distribution, and the LSS clustering, and characterize the uncertainties on their constraints. 

By only assuming homogeneity and isotropy of cosmological perturbations, our method infers the properties of emission sources and the LSS traced by them, as well as the noise in observations from the data covariance $C_{\ell, \nu\nu'}$. We describe the signal with a finite number of emission components, and use the known mapping from the signal rest frame to observed spectral-intensity maps to simultaneously constrain the LSS as well as the SED and redshift evolution of each component. Our method finds the components in a data-driven manner that does not require prior information on the SED of each component or the noise in the data. In that sense, it is similar in spirit to the Spectral Matching Independent Component Analysis \citep[SMICA;][]{2003MNRAS.346.1089D,2008arXiv0803.1814C} algorithm, which  models the covariance of multiple observed CMB frequency maps in terms of a number of components to perform foreground cleaning for CMB data analysis.

While our method provides a new way to probe the LSS from spectral-intensity maps, many studies have analyzed spectral-intensity maps in a different context. The absolute brightness and fluctuations of spectral-intensity maps set constraints on the extragalactic background light (EBL), the aggregate light from all sources of emission across cosmic time. In the optical to near-infrared wavelengths, EBL observations provide important constraints on the background emission behind the resolve sources, such as the diffuse light in the dark matter halos from stripped stars \citep{2012Natur.490..514C,2014Sci...346..732Z,2021ApJ...919...69C,2022ApJ...940..115C}, or the first stars and first galaxies emission from the epoch of reionization and cosmic dawn \citep{2005Natur.438...45K,2011ApJ...742..124M,2012ApJ...753...63K,2015NatCo...6.7945M}. In the far-infrared, the EBL contains crucial information on the high-redshift star formation history and the LSS \citep{2013ApJ...772...77V,2014A&A...570A..98S, 2014A&A...571A..18P,2014A&A...571A..30P}. However, most of these observations are conducted with broadband photometric filters for a higher sensitivity. This makes it challenging to separate signals from different foreground and EBL components, as well as to infer the underlying 3D LSS, as the emission is highly confused along the line of sight. One way to disentangle EBL signals from different redshifts is by cross-correlating EBL maps with tracers of known redshift such as a galaxy catalog \citep{2019ApJ...877..150C, 2022ApJ...925..136C}. However, this only applies to regions where external tracers are available. In contrast, our method can simultaneously extract the 3D LSS and the emission signal from spectral-intensity maps without external information, and we only assume homogeneity and isotropy of the LSS signal, as well as the fact that the emission can be fully described by a finite number of SED components.

Line intensity mapping (LIM) is another emerging technique to probe the 3D LSS from spectral-intensity maps. By mapping a particular spectral line emission, LIM infers the line-of-sight distance of the emission sources from the frequency-redshift relation \citep[e.g. ][]{2017arXiv170909066K, 2022arXiv220615377B}. However, as LIM only analyzes a single spectral line, the majority of  emissions from the full SED have not only been discarded, but also become the continuum \citep[e.g. ][]{2015MNRAS.450.3829Y} or interloper line \citep[e.g.][]{2016ApJ...825..143L,2016ApJ...832..165C,2020ApJ...901..142C} foregrounds in LIM measurements. Our method analyzes the full SED, and thus it is not susceptible to this confusion, and we can also exploit information from emission other than the target spectral line.

\citet{2014arXiv1403.3727D} propose a method to decompose the source SED, redshift dependence, and spatial clustering from spectral-intensity maps by Fourier transforming the spectrum. Despite how exceptional and brilliant their paper is, their method relies on Limber approximation and assumes that all emitting sources can be described by a single SED, which restricts its generalization to larger angular scales and a greater variety of sources in reality. Our data-driven method has the ability to model emission sources with different SEDs in a light cone, enabling more realistic applications than the idealized considerations in \citet{2014arXiv1403.3727D}.

In this work, we provide a proof of concept of our technique with an example setup of a nine-band photometric survey. We generate the mock observed data covariance $C_{\ell,\nu\nu'}$, and use it to perform inference on the underlying signal and noise, and also quantify their uncertainties.

This paper is organized as follows. Sec.~\ref{S:formalism} details the formalism of our technique. Sec.~\ref{S:example_case} describes the observation setup and assumed signal and noise for our example case, and the results of applying our method to this case are presented in Sec.~\ref{S:results}. Sec.~\ref{S:discussion} discusses insights into our method. Sec.~\ref{S:discussion_advantages} highlights the unique advantages of our method. Finally, the conclusion and future outlook are provided in Sec.~\ref{S:conclusion}. Throughout this work, we assume a flat $\Lambda$CDM cosmology with $n_s=0.97$, $\sigma_8=0.82$, $\Omega_m=0.26$, $\Omega_b=0.049$, $\Omega_\Lambda=0.69$, and $h=0.68$, consistent with the measurement from Planck \citep{2016A&A...594A..13P}.


\section{Formalism}\label{S:formalism}
In this section, we describe the formalism for the spectral-intensity signal in a light cone (Sec.~\ref{S:intensity_field}) and its covariance in spherical harmonics space ($C_{\ell,\nu\nu'}$; Sec.~\ref{S:Cl_on_LC}). Then we introduce the parametrization for the signal and noise (Sec.~\ref{S:parametrization}), the likelihood function on parameters (Sec.~\ref{S:likelihood}), and the algorithm for parameter inference (Sec.~\ref{S:parameter_inference}). Here, we only present the formalism and the method synoptically, and provide more detailed derivations in the Appendix.

\subsection{Intensity Field}\label{S:intensity_field}
With the spectral-intensity maps observed in a set of frequencies, we can express the specific intensity $\nu I_\nu$ at the observed frequency $\nu$ and angular position $\hat{n}$ as the integrated emission from all sources in the 3D observing light cone,
\begin{equation}\label{E:nuInu}
\nu I_\nu(\nu,\hat{n})=\int d\chi\int dL\,\Phi(L,\chi,\hat{n})D_A^2(\chi)
\frac{\nu_{\rm rf}L_\nu(\nu_{\rm rf})}{4\pi D_L^2(\chi)},
\end{equation}
where $\chi$ is the co-moving distance, $D_L$ is the luminosity distance, and $D_A$ is the co-moving angular diameter distance, which equals the co-moving distance in a flat universe. $\Phi (L,\chi,\hat{n})=dn(\chi,\hat{n})/dL$ is the luminosity function, defined as the co-moving number density per unit luminosity $L$. Here, $L$ is the total luminosity of a source integrated over its SED, and $L_\nu(\nu_{\rm rf}) \equiv dL/d\nu_{\rm rf}$ is the specific luminosity at the rest-frame frequency\footnote{Throughout this manuscript, $\nu$ is referred to as the observed frequency, and the rest-frame frequency is denoted by $\nu_{\rm rf}$.} $\nu_{\rm rf}=(1+z)\nu$, where $z$ is the redshift of the source\footnote{We use redshift $z$ and co-moving distance $\chi$ interchangeably to describe the line-of-sight distance.}.

Assuming all emitting sources can be classified into $i=1$ to $N_c$ ``components'' of sources, where all sources in each component share the same normalized SED $L^i_\nu(\nu_{\rm rf})/L^i$, and their luminosity function is $\Phi^i(L,\chi,\hat{n})$. The total specific intensity is the sum of emission from all components
\begin{equation}\label{E:nuInui_SMA}
\begin{split}
\nu I_\nu(\nu,\hat{n})&=\sum_{i=1}^{N_c}\int d\chi\int dL^i\,\Phi^i(L^i,\chi,\hat{n})D_A^2(\chi)\frac{\nu_{\rm rf}L^i_\nu(\nu_{\rm rf})}{4\pi D_L^2(\chi)}\\
&=\sum_{i=1}^{N_c}\int d\chi\, S^i(\nu_{\rm rf})M^i_0(\chi,\hat{n})A(\chi),
\end{split}
\end{equation}
where
\begin{align}\label{E:SMP_def}
S^i(\nu_{\rm rf})& \equiv \frac{\nu_{\rm rf}L^i_\nu(\nu_{\rm rf})}{L^i}=\frac{(1+z)\nu L^i_\nu((1+z)\nu)}{L^i},\\
M_0^i(\chi,\hat{n})&\equiv\int dL\,L\Phi^i(L,\chi,\hat{n}),\\
A(\chi)&\equiv\frac{D_A^2(\chi)}{4\pi D_L^2(\chi)}
\end{align}
are normalized SED and luminosity density for component $i$ sources, and the remaining redshift-dependent factors in the integration, respectively. If the intensity field is measured through a filter with frequency response function $R(\nu)$, $S^i$ will be defined as
\begin{equation}
\begin{split}
S^i(\nu_{\rm rf}) &= \frac{1}{\int d\nu R(\nu)}\int d\nu\,R(\nu)\frac{(1+z)\nu L^i_\nu((1+z)\nu)}{L^i}\\
&=\frac{1}{\int d\nu R(\nu)}\int d\nu_{\rm rf}\,R\left(\frac{\nu_{\rm rf}}{(1+z)}\right)\frac{\nu_{\rm rf} L^i_\nu(\nu_{\rm rf})}{L^i}.
\end{split}
\end{equation}
If some sources are masked in the spectral-intensity maps, $M_0^i$ becomes the integration over unmasked sources. See Sec.~\ref{S:number_of_components} for further discussions on the realistic number of components $N_c$.

Note that although in Eq.~\ref{E:nuInu} we describe the intensity field using point-source emitters, our method only relies on the luminosity density $M_0^i$, and therefore we can easily incorporate extended emission to account for diffuse components in the EBL.

\subsection{Angular Power Spectrum on a Light Cone}\label{S:Cl_on_LC}
We measure the information from spectral-intensity maps in covariance space: the auto and cross angular power spectra, $C_{\ell,\nu\nu'}$, between all combinations of frequency bands $\{\nu, \nu'\}$. Owing to the isotropy of the emission field, the set of angular power spectra $C_{\ell,\nu\nu'}$ are a lossless representation of the real-space data covariance. It captures the full information from the dataset at the two-point level \citep{2013acna.conf...87W}. Our inference based on $C_{\ell, \nu\nu'}$ is therefore optimal up to two-point statistics. On large scales, fluctuations can be fully described by a Gaussian probability distribution, and thus the two-point statistics capture the full information from the data. 

The clustering angular power spectrum can be expressed by (see Appendix~\ref{A:power_spectrum} for derivations of equations presented in this section)
\begin{equation}\label{E:Cl_jljl}
\begin{split}
C_{\ell,\nu\nu'}^{\rm clus}&= \sum_{i=1}^{N_c}\int d\chi\, S^i(\nu_{\rm rf})M^i(\chi)A(\chi)\\
&\quad\cdot\sum_{{i'}=1}^{N_c}\int d\chi'\, S^{i'}(\nu'_{\rm rf})M^{i'}(\chi')A(\chi')\\
&\quad\cdot\int \frac{dk}{k}\, \frac{2}{\pi}k^3P(k)G(\chi)j_\ell(k\chi)G(\chi')j_\ell(k\chi'),
\end{split}
\end{equation}
where $P(k)$ is the matter power spectrum at present time, $G(\chi)$ is the linear growth rate, $j_\ell$ is the spherical Bessel function, and $M^i(\chi) \equiv M^i_0(\chi)b^i(\chi)$ is the bias-weighted luminosity density, where $b^i(\chi)$ is the large-scale bias factor. Here, we consider the large-scale linear regime, where the power spectrum transfer function reduces to a scale-independent growth factor $G(\chi)$, and the bias factor $b^i(\chi)$ also has no scale dependence, and we use the linear matter power spectrum for $P(k)$.

We ignore the redshift space distortion (RSD) effect in this work. In reality, the RSD effect results in an additional term to the angular power spectrum $C_{\ell,\nu\nu’}$ that has a different dependence on the luminosity density $M_0^i(\chi)$ and bias $b^i(\chi)$. Therefore, by jointly fitting the RSD and the isotropic clustering term  (Eq.~\ref{E:Cl_jljl}), we can break the degeneracy between $M_0^i(\chi)$ and $b^i(\chi)$. We leave the detailed analysis of RSD to future work.

With an observation of $N_\nu$ spectral bands and angular power spectra in $N_\ell$ bins, we can write all auto/cross spectra at each $\ell$ bin into an $N_\nu\times N_\nu$ matrix,
\begin{equation}\label{E:Cl_clus_matrix}
\mathbf{C}^{\rm clus}_\ell = \mathbf{B}_\ell\mathbf{P}\mathbf{B}_\ell^T,
\end{equation}
where $\mathbf{P}$ is an $N_k\times N_k$ diagonal matrix with its elements being the binned matter power spectrum $P(k)$, and $\mathbf{B}_\ell$ captures all other terms in Eq.~\ref{E:Cl_jljl}.
In practice, the total power spectrum $\mathbf{C}_\ell$ also includes a noise term, $\mathbf{N}_\ell$,
\begin{equation}\label{E:Cl}
\mathbf{C}_\ell = \mathbf{C}^{\rm clus}_\ell+\mathbf{N}_\ell=\mathbf{B}_\ell\mathbf{P}\mathbf{B}_\ell^T+\mathbf{N}_\ell.
\end{equation}
$\mathbf{N}_\ell$ accounts for instrumental noise, foreground contamination, and the Poisson noise from sources. Finally, observations will have stochastic fluctuations, and thus the angular power spectra of the observed data, $\mathbf{C}^d_\ell$, is a random sample from a Wishart distribution with $n_\ell$ degree of freedom, where $n_\ell$ is the number of modes in that $\ell$ bin, and the scale matrix is given by our modeled power spectrum $\mathbf{C}_\ell$ (Eq.~\ref{E:Cl}).

\subsection{Parametrization}\label{S:parametrization}
We parametrize the normalized SED $S^i(\nu_{\rm rf})$, bias-weighted luminosity density $M^i(\chi)$, and the power spectrum $P(k)$ in Eq.~\ref{E:Cl_jljl} to express the clustering power spectra $C_{\ell,\nu\nu'}^{\rm clus}$. We assume $A(\chi)$ and $G(\chi)$ in Eq.~\ref{E:Cl_jljl} are known from the standard cosmological model, while we note that they can be set to any arbitrary function as the parameter dependence is solely captured by the geometry, i.e., the projection law from 3D emission field to the 2D map covariance, regardless of the underlying cosmological model.

We define basis function sets $\left\{\hat{S}(\nu_{\rm rf})\right\}$ and $\left\{\hat{M}(\chi)\right\}$ to linearly expand $S^i(\nu_{\rm rf})$ and $M^i(\chi)$ with $N_s$ and $N_m$ number of basis elements, respectively,
\begin{equation}\label{E:S}
S^i(\nu_{\rm rf}) = \sum_{m=1}^{N_s} c^i_{S,m}\,\hat{S}_m(\nu_{\rm rf}),
\end{equation}
\begin{equation}\label{E:M}
M^i(\chi) = \sum_{n=1}^{N_m} c^i_{M,n}\,\hat{M}_n(\chi),
\end{equation}
and their coefficients, $c^i_{S,m}$ and $c^i_{M,n}$, are the free parameters to be fitted with data. For the 3D power spectrum $P(k)$, the $N_k$ diagonal elements in the matrix $\mathbf{P}$ are our parameters of interest, which represent the averaged band power in each $k$ bin (see Appendix~\ref{A:parametrization} for more implementation details).

The noise matrix $\mathbf{N}_\ell$ has $N_\nu (N_\nu+1)/2$ free parameters for each $\ell$ mode, the same as the degree of freedom in the data $\mathbf{C}_\ell$. Therefore, without  constraints on $\mathbf{N}_\ell$, the noise matrix alone will overfit the data and leave no constraining power for $\mathbf{C}_\ell^{\rm clus}$. Nevertheless, most of the noise sources in reality can be well described by only a few parameters. For example, the instrumental noise usually has negligible correlations between frequency bands, and thus their noise matrices can be mostly diagonal; the Poisson noise from sources is scale-independent, and we show that it can be fully characterized by a redshift-dependent Poisson-to-clustering ratio function (Appendix~\ref{A:Poisson}).  In this work, we consider the noise is uncorrelated between frequency bands, and thus the noise matrix only has $N_\nu$ free parameters as its diagonal elements for each $\ell$ mode. More realistic noise models will be investigated in future work.

In summary, with the observed auto and cross power spectra at $N_\ell$ multipole modes, the total number of data points are $n_d = N_\ell N_\nu(N_\nu+1)/2$, and we fit the data with a set of parameters $\mathbf{\Theta}$ that consists of
\begin{itemize}
    \item $N_c\times N_s$ coefficients $c^i_{S,m}$ for $S^i(\nu_{\rm rf})$
    \item $N_c\times N_m$ coefficients $c^i_{M,n}$ for $M^i(\chi)$
    \item $N_k$ bins of $z=0$ power spectrum $P(k_j)$
    \item $N_\ell\times N_\nu$ noise power spectrum bins $\mathbf{N}_{\ell,\nu\nu}$
\end{itemize}
where $i\in[1,N_c]$, $m\in[1,N_s]$, $n\in[1,N_m]$,  $j\in[1,N_k]$, $\ell\in[1,N_\ell]$, and $\nu\in[1,N_\nu]$ are the indices for source component, $S^i(\nu_{\rm rf})$ basis, $M^i(\chi)$ basis, $k$ bin, $\ell$ bins, and observed frequency bands, respectively. This gives a total number of parameters $n_\theta = N_c(N_s+N_m) + N_k + N_\ell N_\nu$. With the fixed $S^i$ and $M^i$ basis sets and the $\ell$ and $k$ binning, the number of parameters and data ($n_\theta$ and $n_d$) increases with $N_\nu$ and $N_\nu^2$, respectively. This guarantees that we can always get a sufficient degree of freedom from the data to fit all desired model parameters by increasing the number of observing bands. 

\subsection{Likelihood Function}\label{S:likelihood}
Given the observed power spectra $\{\mathbf{C}^d_\ell\}$ in $N_\ell$ multipole bins, we can constrain the parameter set $\mathbf{\Theta}$ using a Bayesian framework. The posterior probability distribution $p$ is
\begin{equation}
p\left ( \mathbf{\Theta}|\{\mathbf{C}^d_\ell\}\right ) \propto \mathcal{L}\left ( \{\mathbf{C}^d_\ell\}|\mathbf{\Theta}\right ) \pi\left ( \mathbf{\Theta}\right ),
\end{equation}
where $\mathcal{L}$ and $\pi$ are the likelihood and prior, respectively. As each $\ell$ mode is independent, the log-likelihood function is the sum of normal distributions for each mode:
\begin{equation}\label{E:likelihood}
\begin{split}
{\rm log} \,\mathcal{L}\left ( \{\mathbf{C}^d_\ell\}|\mathbf{\Theta}\right )&=
-\frac{1}{2}\sum_\ell n_\ell\, {\rm log}\,\mathcal{N}\left ( \mathbf{C}^d_\ell,\mathbf{C}_\ell \right )\\
&=-\frac{1}{2}\sum_\ell n_\ell\left [\right. {\rm Tr}\left ( \mathbf{C}^d_\ell \mathbf{C}_\ell^{-1}\left ( \mathbf{\Theta} \right ) \right )\\
&+{\rm log\,det}\left ( \mathbf{C}_\ell\left ( \mathbf{\Theta} \right ) \right ) +N_\nu {\rm log} \left(2\pi\right)\left.\right ],
\end{split}
\end{equation}
where $n_\ell$ is the number of modes in each $\ell$ bin (Eq.~\ref{E:number_of_modes}).

The overall amplitudes between $S^i$, $M^i$, and $P$ are degenerate in $C_{\ell,\nu\nu'}^{\rm clus}$ (Eq.~\ref{E:Cl_jljl}),  and thus we introduce a regularization term to our prior ($\pi_{\rm reg}$) to break the degeneracy (Appendix~\ref{A:regularization}). In the cases of multiple source components ($N_c>1$), there is a strict symmetry under exchange/permutation of the components, i.e., swapping $S^i$ and $M^i$ between two components results in the same clustering power spectra. Nevertheless, this degeneracy gives multiple identical and separate peaks in the likelihood function, and all those peaks are equally valid, since they are just different by the inferred order of components. This is contrary to the $S^i$, $M^i$, and $P$ normalization degeneracy, which gives a continuous flat hyper-surface of maximum likelihood in the parameter space. The discrete degeneracy of components can be removed exactly by defining any unique ordering for the components, but even without this process, we can still derive the parameter constraints by the likelihood function around one of the solutions. 

\subsection{Parameter Inference}\label{S:parameter_inference}
With a set of angular power spectra from data $\{\mathbf{C}_\ell^d\}$, we use the Newton-Raphson method to find the set of parameters $\Theta_{\rm max}$ at the maximum likelihood
\begin{equation}
\Theta_{\rm max} = \max_{\Theta} {\rm ln}\,\mathcal{L}.
\end{equation}
Then we estimate the parameter constraints with the Fisher matrix at $\Theta_{\rm max}$. The Fisher matrix is given by
\begin{equation}\label{E:Fisher}
\begin{split}
\mathbf{F}_{\alpha\beta} &= -\left \langle \frac{\partial^2 {\rm log}\,\mathcal{L}}{\partial\theta_\alpha\partial\theta_\beta} \right \rangle=
\sum_\ell \mathbf{F}_{\ell,\alpha\beta}\\
&=\frac{1}{2}\sum_\ell n_\ell {\rm Tr}\left ( \mathbf{C}_\ell^{-1}\frac{\partial\mathbf{C_\ell}}{\partial\theta_\alpha}\mathbf{C}_\ell^{-1}\frac{\partial\mathbf{C_\ell}}{\partial\theta_\beta} \right ),
\end{split}
\end{equation}
and the parameter covariance is the inverse of the Fisher matrix (see Appendices \ref{A:Newtons} and \ref{A:Fisher} for detailed derivations and implementations on the Newton-Raphson method and Fisher matrix, respectively). However, due to the degeneracy of $S^i$, $M^i$, and $P$, the Fisher matrix is singular and cannot be inverted to obtain the covariance matrix. Therefore, we add the regularization prior $\pi_{\rm reg}$ to the likelihood before inverting the Fisher matrix, which gives our covariance matrix estimator
\begin{equation}\label{E:Fisher_inv}
\hat{\mathbf{F}}^{-1}=\left ( \mathbf{F}_{\rm reg}+\mathbf{F} \right )^{-1}.
\end{equation}
See Appendix~\ref{A:regularization} for the expression of $\mathbf{F}_{\rm reg}$.

The combination of Newton-Raphson method and the Fisher matrix formalism give us fast and accurate posterior approximations. We also validate our results on parameter inference with the Markov Chain Monte Carlo (MCMC). Appendix~\ref{A:MCMC} describes the implementation details of our MCMC sampling.

\section{Example Case}\label{S:example_case}
We will demonstrate our algorithm with some simple example cases. We consider a mock survey taking spectral-intensity maps in several spectral bands, assume a model of the signal (source SED $S^i(\nu_{\rm rf})$, luminosity density $M^i(\chi)$, and power spectrum $P(k)$) and the noise, and calculate the auto and cross angular power spectrum $C_{\ell,\nu\nu'}$ from this observation. Then we apply our inference algorithms (Sec.~\ref{S:parameter_inference}) to derive constraints on signal and noise. This section describes the assumed survey setup, signal and noise model, and our choice of parameters for the fiducial case.

\subsection{Survey Setup}

\begin{figure}[ht!]
\begin{center}
\includegraphics[width=\linewidth]{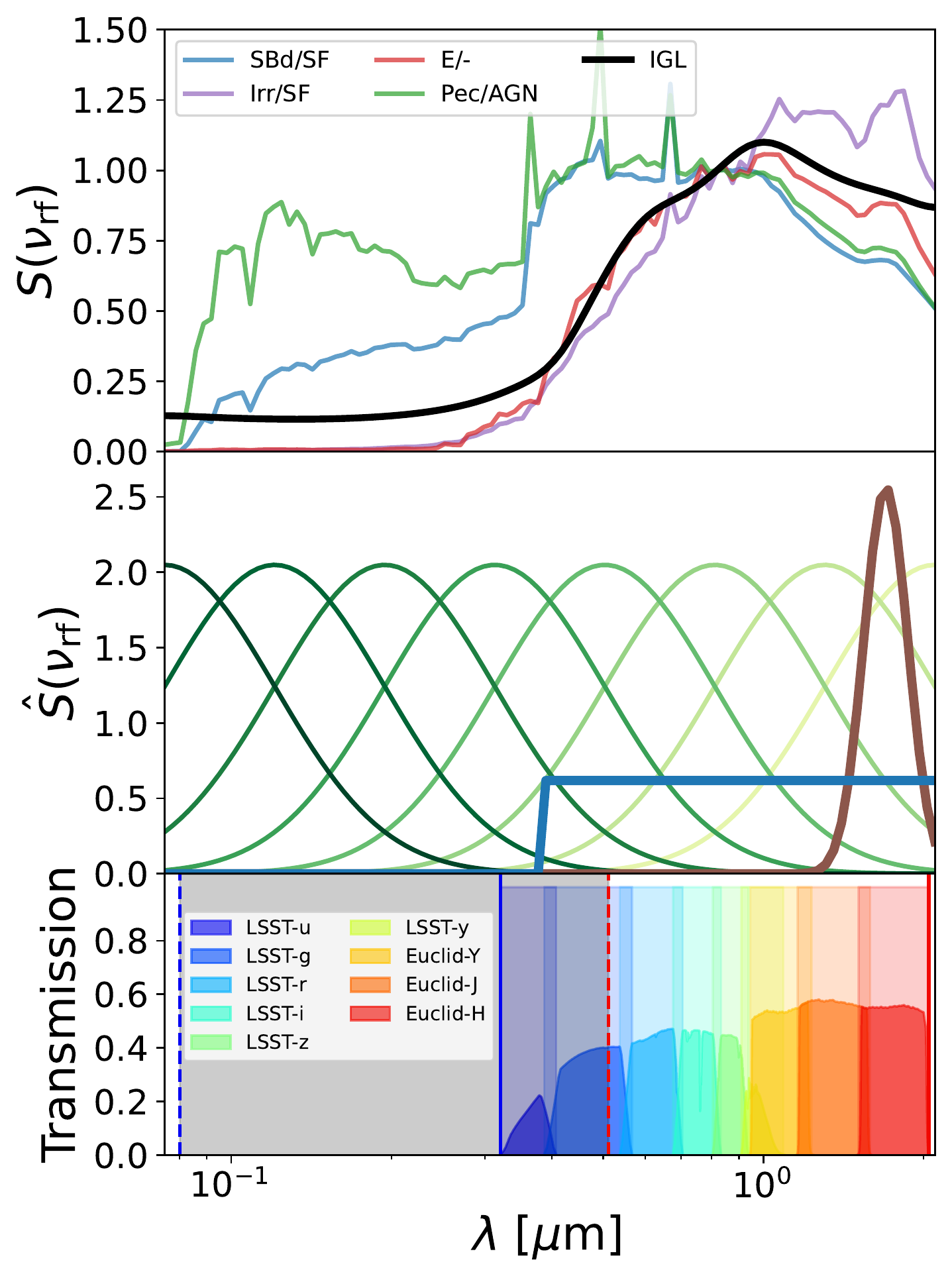}
\caption{\label{F:model_S} Top: SED of the H12 IGL model at $z=0$ (black). Color lines show SEDs of four local galaxies with different morphology and BPT diagram classes from \citet{2014ApJS..212...18B} (morphology class/BPT class for emission line galaxies): NGC 0337 -- SBd/star-forming (SBd/SF; blue), NGC 2388 -- Irregular/star-forming (Irr/SF; purple), NGC 4365 -- elliptical/BPT classification not defined (E/--; red), and UGCA 208 -- Peculiar/active galactic neclei (Pec/AGN, green). We normalize all SEDs at $0.8$ $\mu$m. Middle: the SED basis set $\hat{S}_{m}$ considered in this work. We use eight Gaussian functions ($\hat{S}_{G,m}$) equally spaced in logarithmic wavelength to expand the continuum spectrum (green curves), another narrower Gaussian centered at $1.6$ $\mu$m ($\hat{S}_{1.6}$) to model the $1.6$ $\mu$m bump (brown), and a Heaviside step function at $0.4$ $\mu$m ($\hat{S}_{0.4}$) to represent the 4000 $\AA$ break. Bottom: filter transmission profile of the six Rubin Observatory LSST and three Euclid photometric bands (dark-shaded color regions). Here, we use top-hat filters with similar wavelength coverage for our fiducial case (light-shaded color regions). The gray band shows the rest-frame wavelengths probed by our observing wavelength range at $z=3$, which is the maximum redshift considered in our model. The boundaries of the gray band (blue- and red-dashed lines) are the wavelengths of blue and red solid lines redshifted to $z=3$.}
\end{center}
\end{figure}

We consider our spectral-intensity maps observed from nine broadband observations corresponding to two upcoming photometric surveys: Rubin Observatory LSST \citep{2009arXiv0912.0201L} and Euclid \citep{2011arXiv1110.3193L}.  LSST will cover a total area of $18,000$ deg$^2$ in six optical bands ($u$, $g$, $r$, $i$, $z$); the Euclid NISP instrument will map $15,000$ deg$^2$ in three near-infrared bands ($Y$, $J$, $H$)\footnote{We do not consider the Euclid VIS band as it  overlaps with the LSST wavelengths.}. For simplicity, we use a top-hat frequency response function with similar wavelength coverage as the LSST and Euclid filters (Fig.~\ref{F:model_S}).
We assume a total survey area of $11,000$ deg$^2$ ($f_{\rm sky}=0.27$), which is the size of the LSST-Euclid overlapping area if LSST extends their survey to the low declination area proposed by \citet{2017ApJS..233...21R}.  Note that equivalently, the combination of the Nancy Grace Roman Space Telescope High Latitude survey \citep{2015arXiv150303757S} and LSST would cover $2000$ deg$^2$ with 10 bands. We use 30 logarithmically spaced $\ell$ bins within $10^1\leq \ell \leq 3\times10^4$, corresponding to tens of arcsecond to tens of degree scales.

\subsection{Signal Model}\label{S:signal_model}
We consider only a single source component in our fiducial case ($N_c=1$), and show an example with two components in Sec.~\ref{S:discussion_2cls}. While it is unrealistic to assume all emission sources have the same SED, this simplification is reasonable for modeling the large-scale signals: as the power spectrum at a mode $\ell$ corresponds to the real-space correlation of the mean fluctuations in a region with angular size $\sim \ell^{-1}$, the signal on large scales (low $\ell$) can be well described by the mean SED of all emission sources. \citet[][hereafter H12]{2012ApJ...752..113H} built the galaxy luminosity function across redshift based on several galaxy counts observations, and used it to model the  integrated galactic light (IGL), the aggregate emission from all galaxies across redshift. The top panel of Fig.~\ref{F:model_S} shows the SED of the IGL from $z=0$ sources from the H12 model, which is equivalent to the mean SED of all local galaxies. For comparison, we also show the SEDs of four local galaxies from \citet{2014ApJS..212...18B} with different morphology and BPT diagram classes.

\begin{figure}[ht!]
\begin{center}
\includegraphics[width=\linewidth]{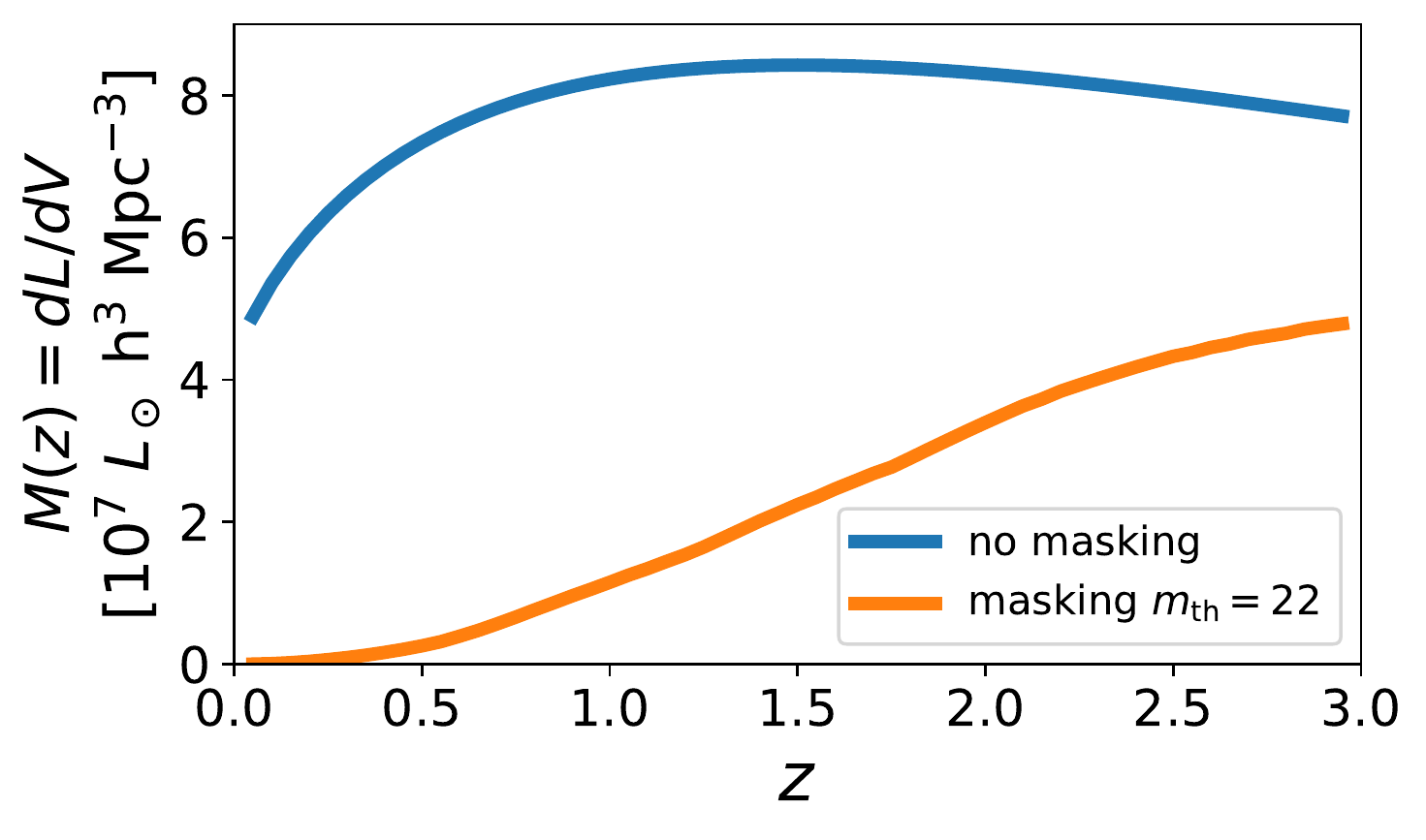}
\caption{\label{F:model_M} IGL luminosity density from H12 integrated over rest frame $0.15$--$5$ $\mu$m. The blue line denotes the luminosity density from all IGL sources, and the orange line denotes the case of masking galaxies with a magnitude threshold $m_{\rm th}=22$ at $1$ $\mu$m.}
\end{center}
\end{figure}

Fig.~\ref{F:model_M} shows the luminosity density $M_0(\chi)$ from the H12 IGL model. Here, we integrate the total luminosity from rest frame $0.15$--$5$ $\mu$m, and assume $b(\chi)=1$ (so $M(\chi)=M_0(\chi)$). In this work, we consider emission from $0\leq z\leq 3$, 

For our fiducial case, $S(\nu_{\rm rf})$ and $M(\chi)$ are not set to the IGL model from H12; instead, we use linear combinations from our basis functions in order to directly compare the parameter constraints with their input values (see Sec.~\ref{S:fiducial_case}).

\begin{figure}[ht!]
\begin{center}
\includegraphics[width=\linewidth]{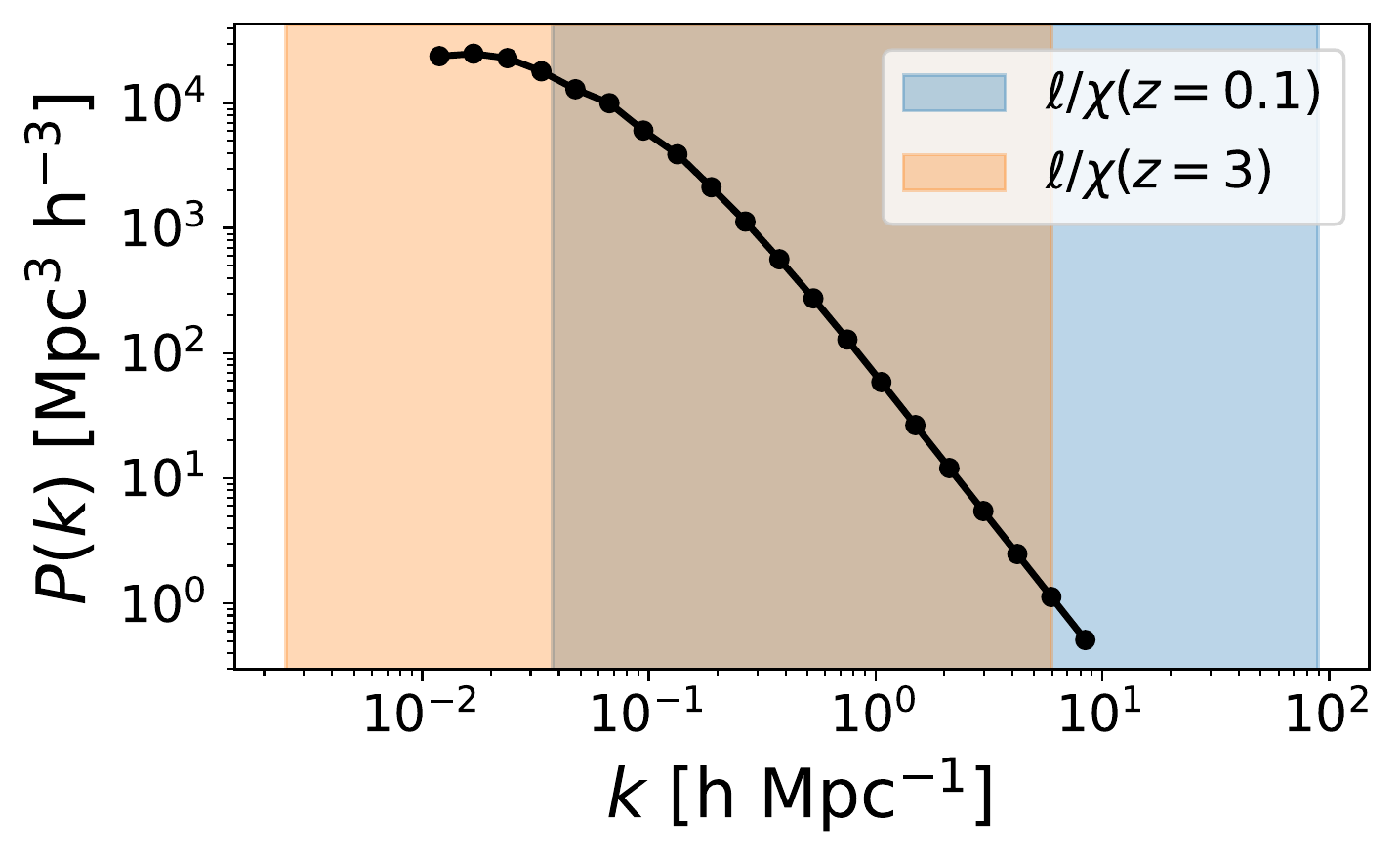}
\caption{\label{F:model_P} The three-dimensional linear matter power spectrum used in our model. We use 20 logarithmically spaced bins in the range of $10^{-2} \leq k \leq 10^{1}$ $h$ Mpc$^{-1}$. Blue and orange shaded regions show the range of transverse $k$ modes corresponding to the range of $\ell$ modes considered in this work ($10^1\leq \ell \leq 3\times10^4$) at redshift $0.1$ and $3$, respectively.}
\end{center}
\end{figure}

For the three-dimensional power spectrum $P(k)$, we use the linear matter power spectrum with 20 logarithmically spaced bins in the range $10^{-2} \leq k \leq 10^{1}$ $h$ Mpc$^{-1}$ (Fig.~\ref{F:model_P}). We consider our model at $0\leq z \leq 3$, and show the transverse $k$ mode range corresponding to our $\ell$ range ($10^1\leq \ell \leq 3\times10^4$) at $z=0.1$ and $z=3$.

\begin{figure}[ht!]
\begin{center}
\includegraphics[width=\linewidth]{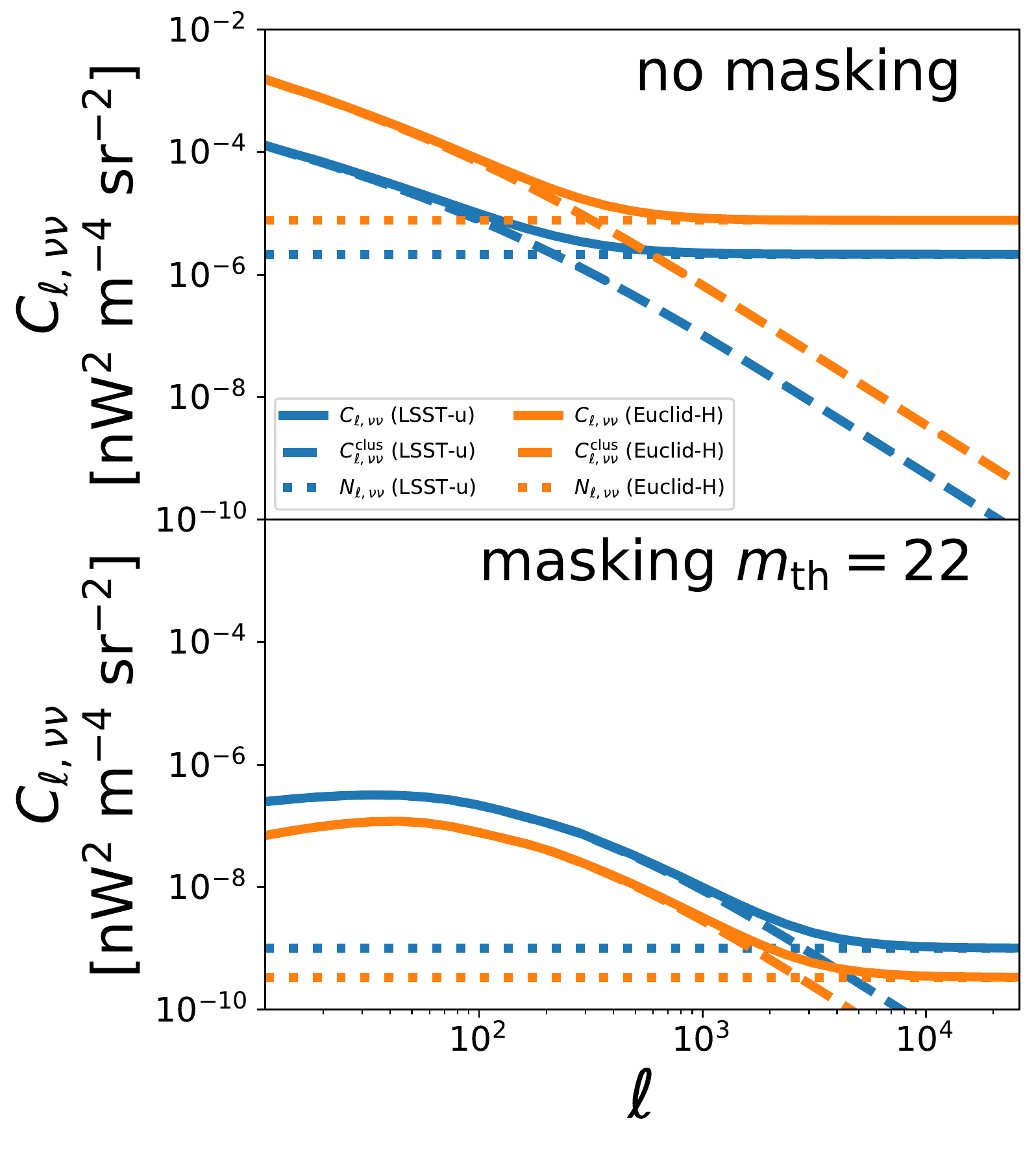}
\caption{\label{F:model_Cl} IGL clustering (dashed) and Poisson noise (dotted) power spectrum from $z=0$--$3$ with the H12 model. We show the auto power spectrum at the shortest (LSST-u; blue) and longest (Euclid-H; orange) wavelength bands considered in this work. Solid lines show the total power spectrum (sum of the clustering and Poisson noise terms). The top panel is the case without masking, and the bottom panel is the case of masking sources with a magnitude threshold $m_{\rm th}=22$ at 1 $\mu$m.}
\end{center}
\end{figure}

For the noise matrix $\mathbf{N}_\ell$, we consider the noise has no scale ($\ell$) dependence, and it is uncorrelated between frequency bands, and therefore $\mathbf{N}_\ell$'s are diagonal matrices and are identical for all $\ell$ modes. We set the noise power spectrum level at each frequency band to the Poisson noise level from the H12 model, although in reality Poisson noise will be strongly correlated across frequencies (see Appendix~\ref{A:Poisson}). We will include this consideration in future work. The clustering signal and Poisson noise power spectrum from the H12 model are shown in Fig.~\ref{F:model_Cl}. 

\begin{figure*}[ht!]
\includegraphics[width=\linewidth]{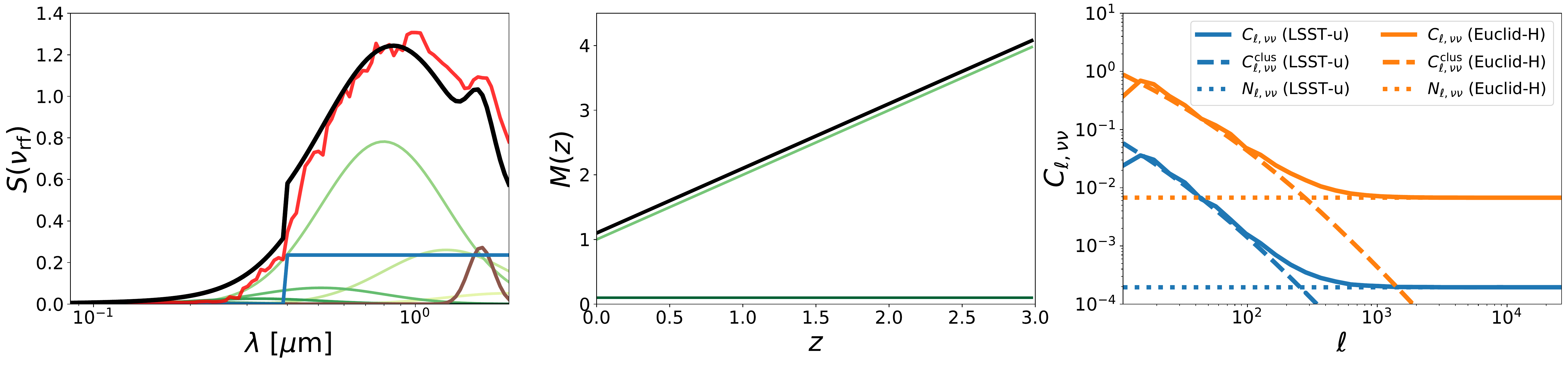}
\caption{\label{F:mock_model} The model of our fiducial case. Left: our fiducial $S(\nu_{\rm rf})$ model (black) built from the linear combination of the basis function set, i.e., $S(\nu_{\rm rf})=\sum_{m=1}^{N_s} c_{S, m}\hat{S}_m(\nu_{\rm rf})$, and each basis component ($c_{S, m}\hat{S}_m(\nu_{\rm rf})$) is shown by the green, blue, and brown curves with the same color coding as in Fig.~\ref{F:model_S}. The amplitudes $c_{S, m}$ are defined such that our $S(\nu_{\rm rf})$ resembles the SED of the elliptical galaxy NGC 4365 \citep[red, ][]{2014ApJS..212...18B}. Middle: our fiducial model of $M(z)$ (black). We use the zeroth and the first-order polynomial basis, i.e., $M(z)=c_{M,0}\hat{M}_0(z)+c_{M,1}\hat{M}_1(z)$, and the dark and light green curves show the two basis components. Right: the angular auto power spectrum of our fiducial model in LSST-u (blue) and Euclid-H (orange) bands. Dashed and dotted lines denote the clustering and noise power spectrum, respectively. For the noise model, we assume white noise with no correlation between frequency bands. The noise power spectrum amplitude is set such that its relative power to the clustering signal is similar to the Poisson noise-to-clustering power spectrum ratio in the H12 model with masking threshold $m_{\rm th}=22$ case (bottom panel of Fig.~\ref{F:model_Cl}). The observed power spectra (solid curves) are the sum of the clustering and noise term with a sample variance following the Wishart distribution (Appendix~\ref{S:Cl_variance}). Note that all parameters here are set with arbitrary normalization, so their values do not represent any physical units.}
\end{figure*}

In practice, it is usually beneficial to mask bright/detected sources in order to better probe the signal from the faint background emission. For our fiducial case, we mask sources brighter than an AB magnitude threshold $m_{\rm th}=22$ at $1$ $\mu$m. The masking threshold for other wavelengths is set with the abundance matching prescription from \citet{2022ApJ...925..136C}. The $M(\chi)$ and the angular power spectrum with masking are also shown in Fig.~\ref{F:model_M} and \ref{F:model_Cl}. The point-source sensitivity required for this masking depth can be achieved by a single LSST exposure (LSST $y$ band $5\sigma$ point-source depth is 22.1 mag\footnote{\url{https://www.lsst.org/scientists/keynumbers}}). We test that this magnitude threshold effectively removes a large portion of bright sources, and further deepening the masking threshold does not improve the clustering-to-Poisson noise ratio significantly. Note that the masked bright sources also contain important information, but since we usually have prior knowledge of their SEDs and distances from photometric/spectroscopic redshift measurements, the optimal way to use these detected sources in the analysis is to incorporate this prior knowledge instead of inferring it blindly with our method. This can be done by, for example, cross-correlating the masked spectral-intensity maps with the 3D distribution of detected galaxies, which will be investigated in future work.

From Fig.~\ref{F:model_Cl}, we can see that masking can increase the clustering-to-Poisson noise ratio, since bright sources have higher weights in the Poisson noise than in the signal. Another benefit of our magnitude-limited masking scheme is to enhance the high-redshift emission in the observed signal (see Fig.~\ref{F:model_M}). This is because with a fixed brightness threshold, the low-redshift population will be masked to a deeper absolute brightness level than the high-redshift sources. This will help improve constraints on the 3D power spectrum $P(k)$ at lower $k$, since for the same observed angular scale, the higher redshift emission corresponds to larger co-moving scales. This effect can be seen in Fig.~\ref{F:model_Cl}, where for the case with masking, the angular power spectrum $C_\ell$ reflects the shape of $P(k)$ at lower-$k$ modes compared to the case without masking.

\subsection{Basis Functions}\label{S:basis}
For the SED basis ($\hat{S}_m$), we use a set of 10 basis functions (see the middle panel of Fig.~\ref{F:model_S}). Eight of them are Gaussian functions used to span the continuum SED:
\begin{equation}
\begin{split}
\hat{S}_{G,m}(\nu_{\rm rf})=&\frac{1}{\sqrt{2\pi \left ( {\rm log}_{10}\sigma_G \right )^2}}\\
&\cdot{\rm exp}\left [ \frac{-\left ( {\rm log}_{10}\nu_{\rm rf}  - {\rm log}_{10}\nu_{G,m}\right )^2}{2\left ( {\rm log}_{10}\sigma_{G}\right )^2} \right ].
\end{split}
\end{equation}
The center frequencies, ${\rm log}_{10}\nu_{G,m}$, are logarithmically spaced in wavelength (frequency) from $0.85$--$1.95$ $\mu$m, and their standard deviation, ${\rm log}_{10}\sigma_{G,m}=0.195$, is the same as their spacing.

Another basis component is a narrow Gaussian peaked at $1.6$ $\mu$m to model the ``$1.6$ $\mu$m bump'' arising from the minimum of $H^{-}$ opacity \citep{1988A&A...193..189J}:
\begin{equation}
\begin{split}
\hat{S}_{1.6}(\nu_{\rm rf})=&\frac{1}{\sqrt{2\pi \left ( {\rm log}_{10}\sigma_{1.6} \right )^2}}\\
&\cdot{\rm exp}\left [ \frac{-\left ( {\rm log}_{10}\nu_{\rm rf}  - {\rm log}_{10}\nu_{1.6}\right )^2}{2\left ( {\rm log}_{10}\sigma_{1.6}\right )^2} \right ],
\end{split}
\end{equation}
where $\nu_{1.6}=c/(1.6\, \mu{\rm m})$, and we set ${\rm log}_{10}\sigma_{1.6}=0.039$. 

The last basis component is used to model the  ``4000 $\AA$ break'', a typical SED feature in early-type galaxies caused by the lack of blue stars and the blanket absorption of high-energy photons from metals \citep{1963AJ.....68..413V}. We use a Heaviside step function $\mathcal{H}$ to describe the 4000 $\AA$ break:
\begin{equation}\label{E:Sbasis_4000}
\hat{S}_{0.4}(\nu_{\rm rf})=\mathcal{H}(\nu_{0.4}),
\end{equation}
where $\nu_{0.4}=c/(4000 \AA)$.

The (bias-weighted) luminosity density is expected to be a smooth function of redshift, so we use polynomials as its basis:
\begin{equation}\label{E:M_basis}
\hat{M}_n(z) = (1+z)^n.
\end{equation}

\subsection{Fiducial Case}\label{S:fiducial_case}
For our fiducial example, instead of using the H12 model presented in Sec.~\ref{S:signal_model}, we build the $S(\nu_{\rm rf})$ and $M(\chi)$ from our basis functions (Fig.~\ref{F:mock_model}). This allows us to directly compare the parameter constraints from our algorithm to the ground truth input values. For $S(\nu_{\rm rf})$, we set the relative amplitude of each basis $\hat{S}$ such that the SED shape of our model resembles the elliptical galaxy NGC 4365 \citep{2014ApJS..212...18B}. For $M(\chi)$ we use only the zeroth and first-order polynomials, and thus our luminosity density is a linear function of redshift. Our model of $P(k)$ is the linear matter power spectrum shown in Fig.~\ref{F:model_P}. For the noise matrix, we consider white noise without cross-channel correlation, so $\mathbf{N}_{\ell}$ matrices are diagonal and are identical for all $\ell$ modes. We set the noise power spectrum such that its amplitude relative to the clustering signal is similar to that of H12 model with $m_{\rm th}=22$ case (bottom panel of Fig.~\ref{F:model_Cl}). 

In summary, our fiducial case has $N_c=1$, $N_s=10$, $N_m=2$, $N_k=20$, $N_\ell=30$, and $N_\nu=9$, which gives the total number of parameters $N_\theta = 302$.

\section{Results}\label{S:results}

This section presents the results of parameter constraints with the fiducial case. We verified that for all cases investigated in this work, the posterior inference derived from the combination of the Newton-Raphson method and the Fisher matrix is consistent with MCMC. Here, we show the results that include sample variance fluctuations in the mock data, and perform inference with MCMC.

\begin{figure}[ht!]
\begin{center}
\includegraphics[width=\linewidth]{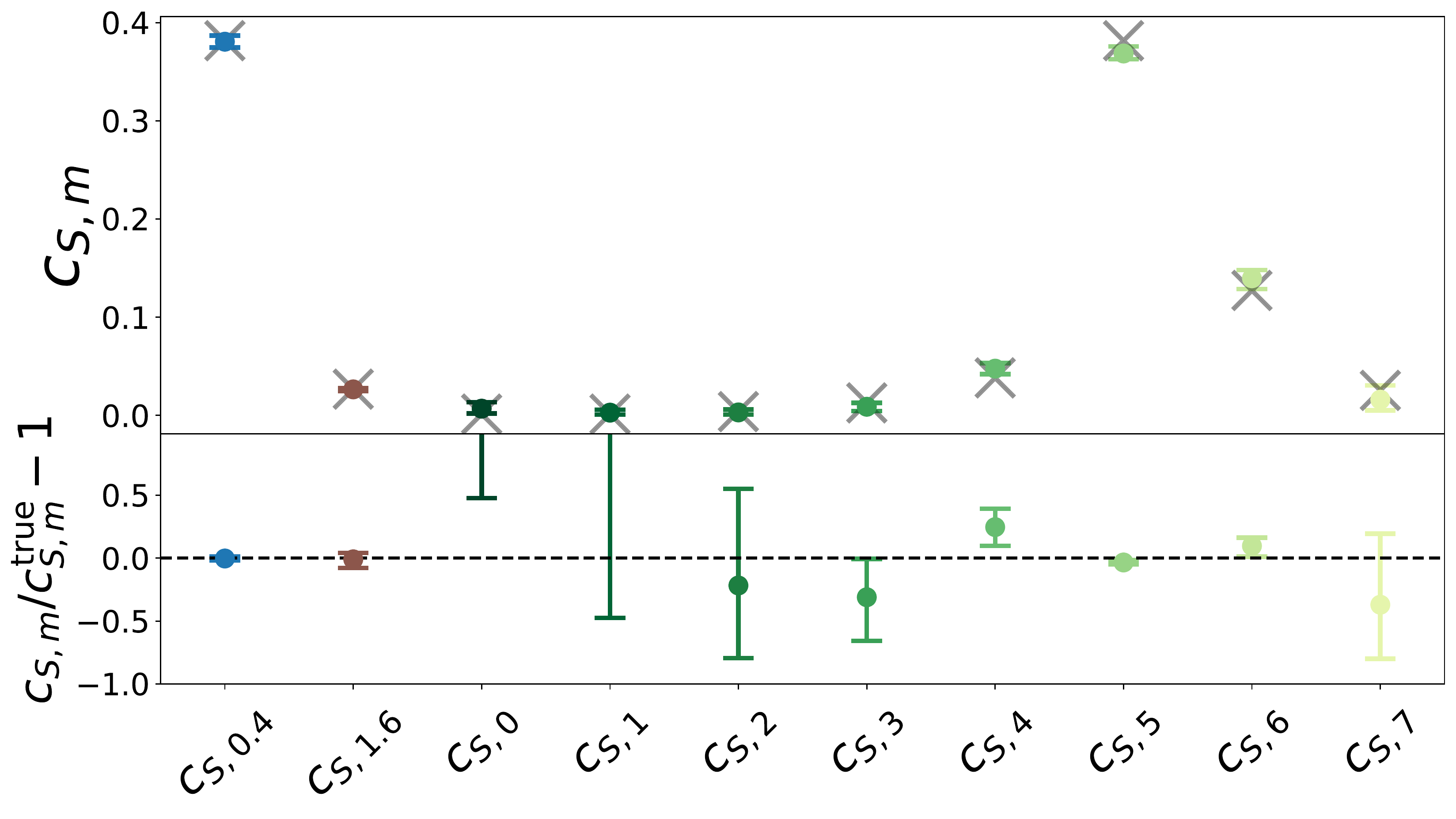}
\includegraphics[width=\linewidth]{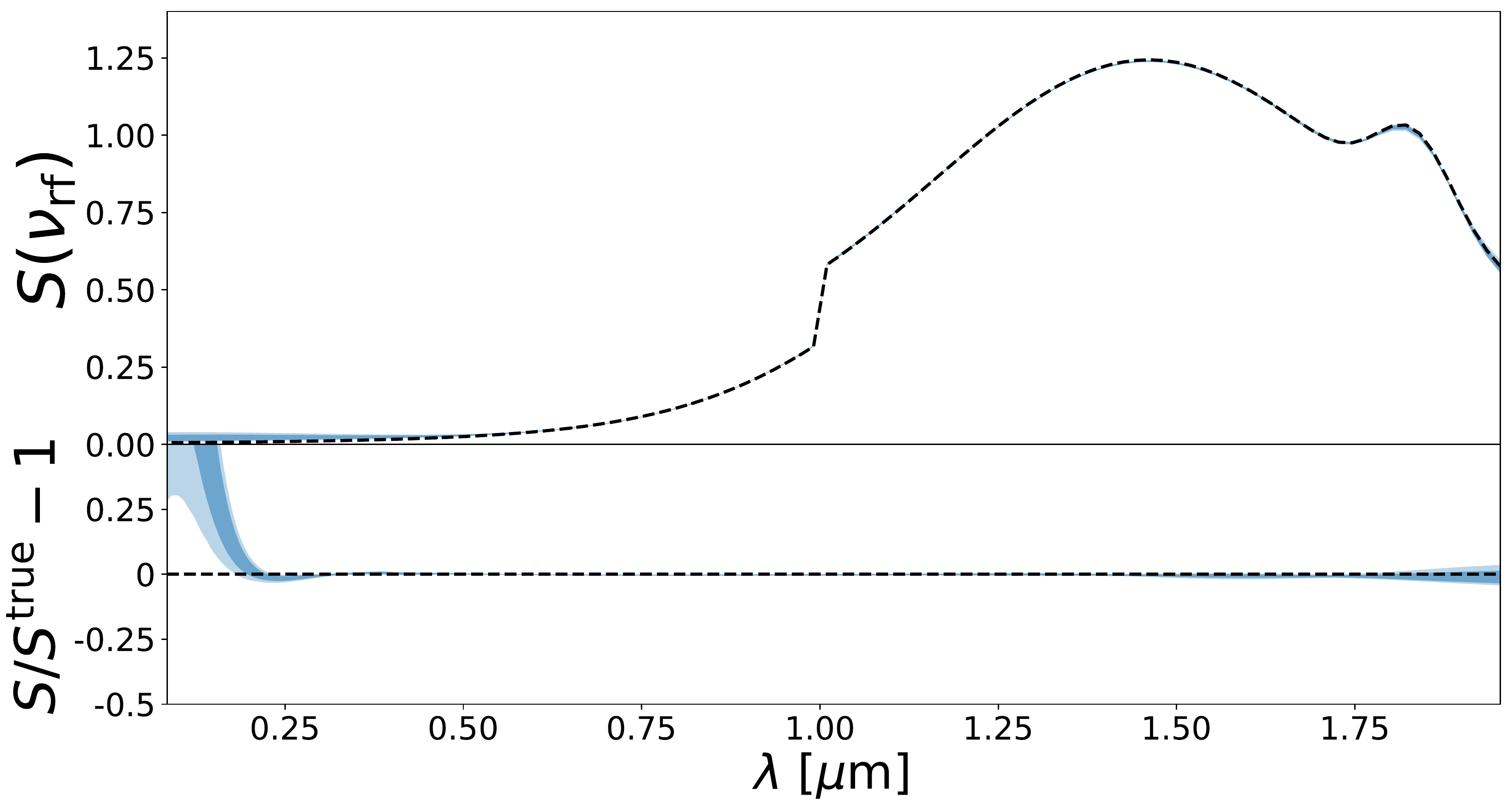}
\caption{\label{F:S} MCMC posterior on $S^i(\nu_{\rm rf})$ with the fiducial case. Top: constraints on the $S^i(\nu_{\rm rf})$ basis function amplitudes, $c_{S,m}$, with the same color coding as in Fig.~\ref{F:model_S} for each basis component. Error bars denote the 68th percentile of the marginalized posterior, and the gray crosses denote the truth values. Bottom: the 68th (dark blue) and 95th (light blue) posterior percentiles of the reconstructed source SED $S^i(\nu_{\rm rf})$, and the black-dashed line denotes the truth input mode. The bottom subpanel in each plot shows the fractional error against the truth values.}
\end{center}
\end{figure}

\begin{figure}[ht!]
\begin{center}
\includegraphics[width=\linewidth]{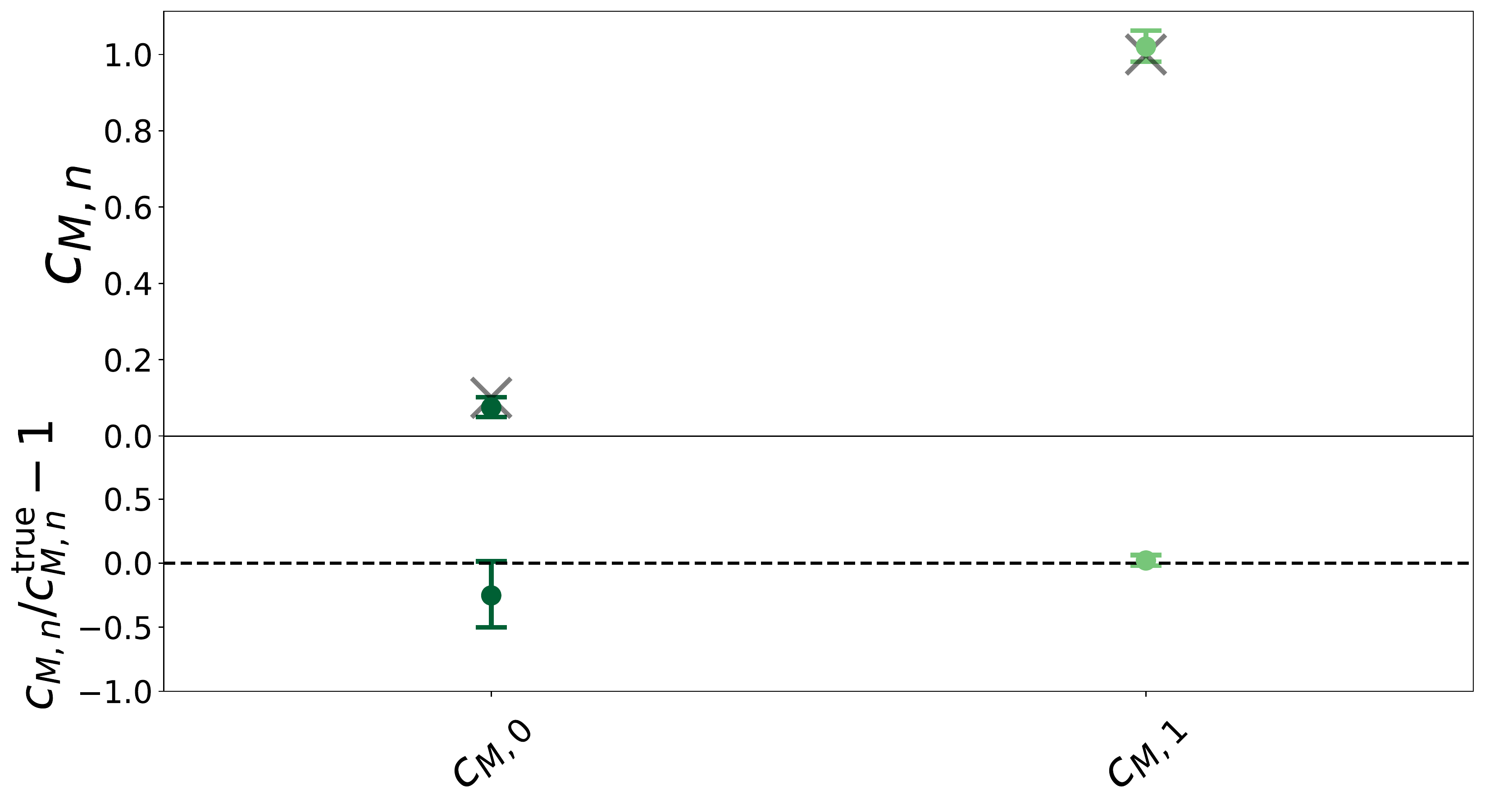}
\includegraphics[width=\linewidth]{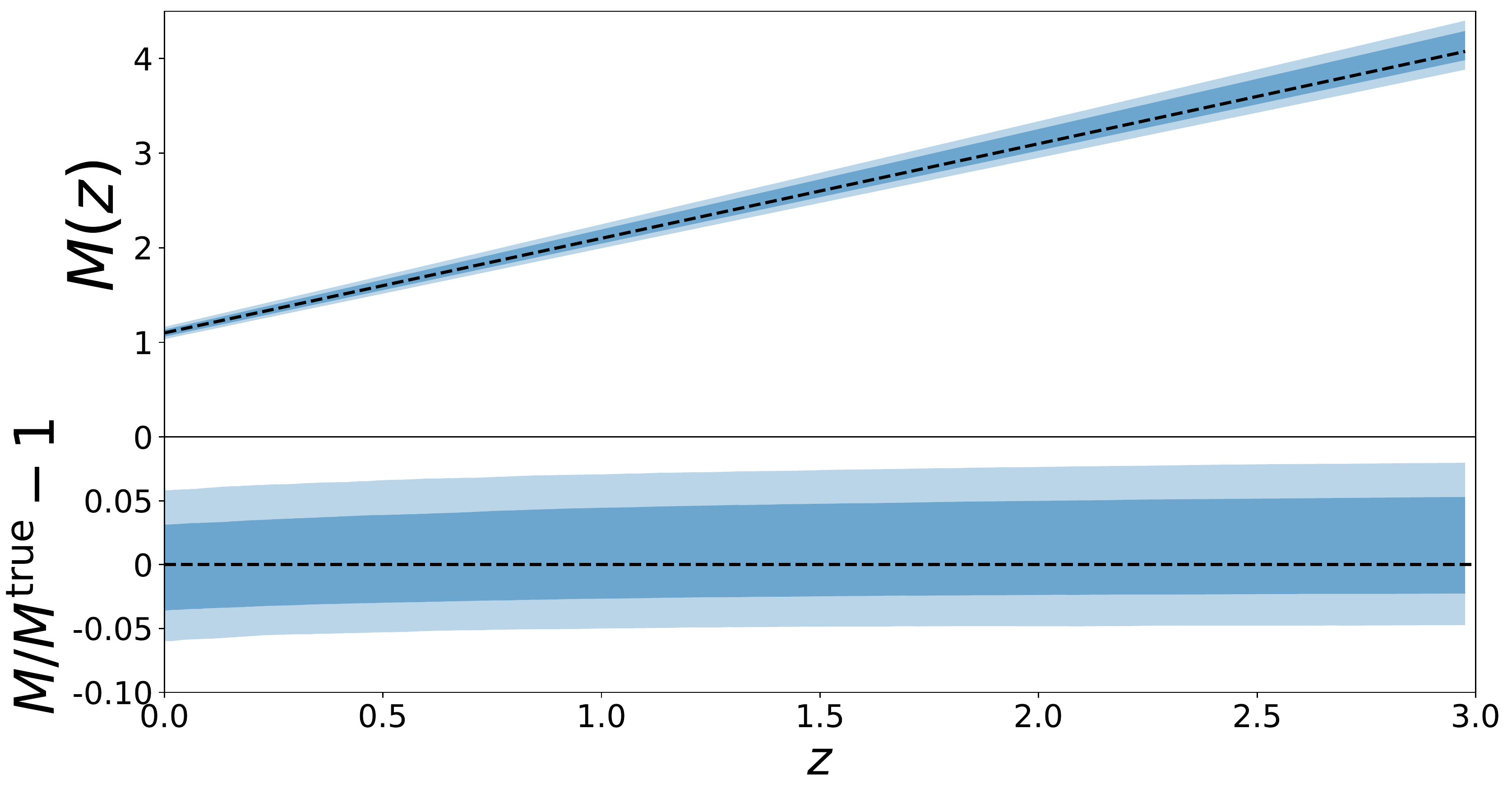}
\caption{\label{F:M} MCMC posterior on $M^i(\chi)$ with the fiducial case. Top: constraints on the $M^i(\chi)$ basis function amplitudes $c_{M,n}$. Error bars indicate the 68th percentile of the marginalized posterior, and the gray crosses denote the truth values. Bottom: the 68th (dark blue) and 95th (light blue) posterior percentiles of the reconstructed $M^i(\chi)$, and the black-dashed line indicates the truth input model. The bottom subpanel in each plot shows the fractional error against the truth values.}
\end{center}
\end{figure}

\begin{figure}[ht!]
\begin{center}
\includegraphics[width=\linewidth]{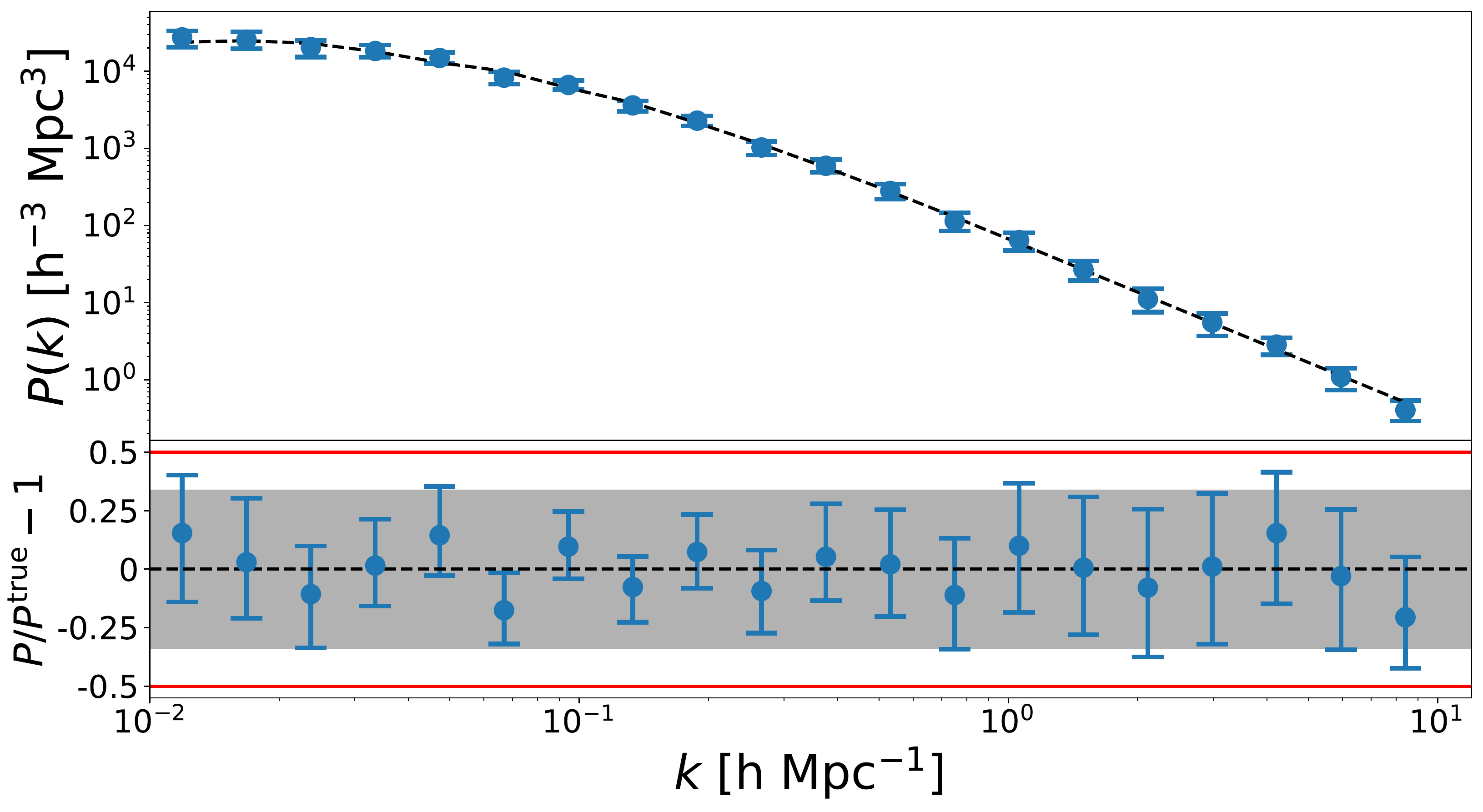}
\caption{\label{F:P} Top: MCMC posterior on $P(k)$ with the fiducial case. Error bars denote the 68th percentile of the marginalized posterior, and the black-dashed line denotes the truth input model. Bottom: fractional error against the truth values. For reference, the red lines mark the prior limit of the power spectrum ($\pm50\%$ of the truth) where the prior probability is set to zero outside this range, and the gray band indicates the 68th percentile of the prior probability distribution.}
\end{center}
\end{figure}

\begin{figure}[ht!]
\begin{center}
\includegraphics[width=\linewidth]{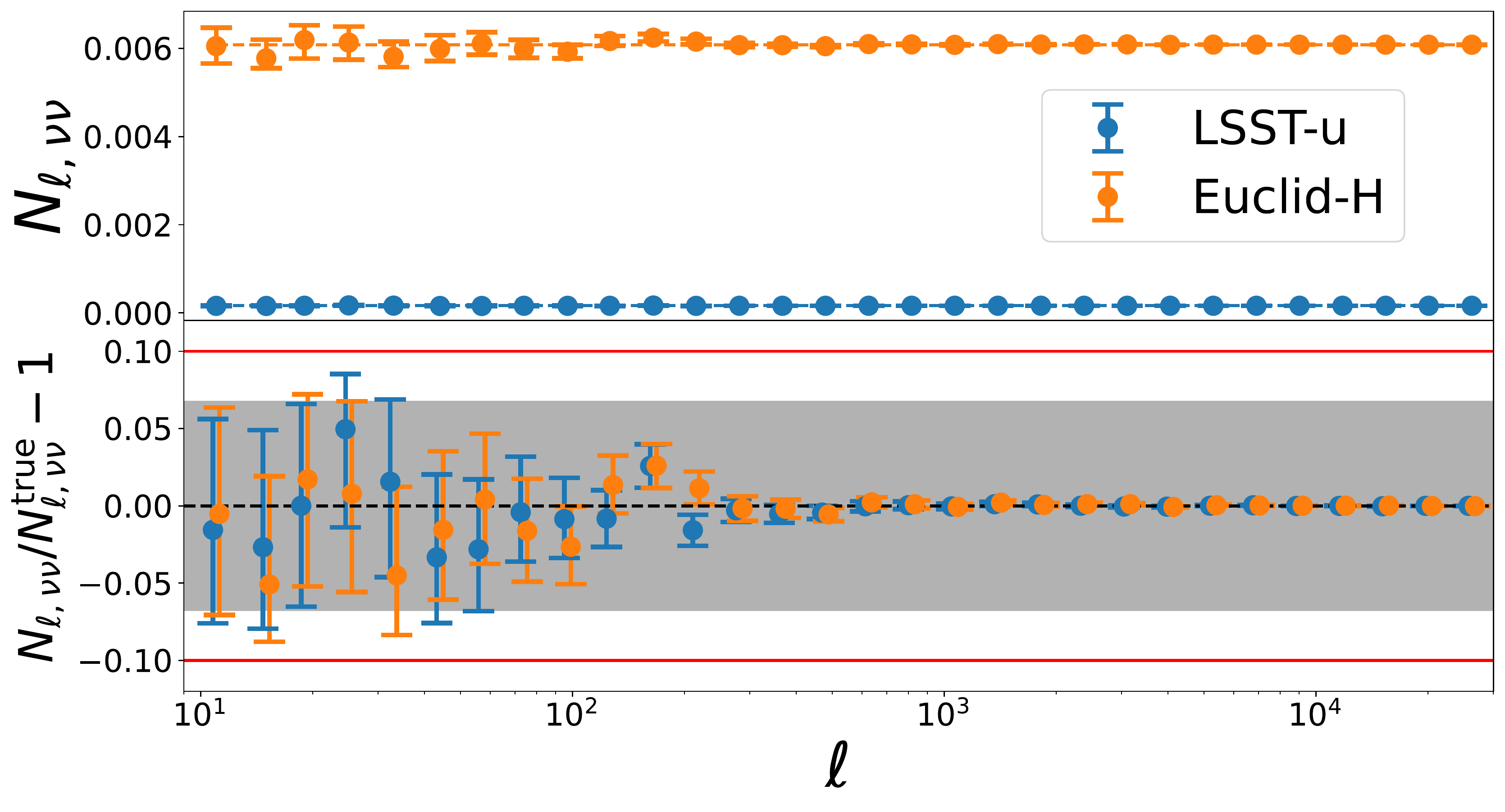}
\caption{\label{F:N} Top: MCMC posterior on $\mathbf{N}_{\ell}$ with the fiducial case in LSST-u and Euclid-H bands. Error bars denote the 68th percentile of the marginalized posterior, and the dashed lines denote the truth input model. Bottom: fractional error against the truth values. For reference, the red lines mark the prior limit of the power spectrum ($\pm10\%$ of the truth) where the prior probability is set to zero outside this range, and the gray band marks the 68th percentile of the prior probability distribution.}
\end{center}
\end{figure}

Fig.~\ref{F:S}, \ref{F:M}, \ref{F:P}, \ref{F:N} show the marginalized posterior on $S^i$, $M^i$, $P$, and $\mathbf{N}_{\ell}$, respectively. With our fiducial setup, the luminosity density $M^i(z)$ can be determined unbiasedly with $\sim 5\%$ uncertainty; the SED of emission sources, $S^i(\nu_{\rm rf})$, can also be inferred with percent-level of errors except for $\lambda_{\rm rf}\gtrsim 0.2$ $\mu$m. The rest-frame SED at shorter wavelengths is less constrained since it is intrinsically fainter in our model, and also it can only be probed by high-redshift signal given our observing bands. The amplitude of the 4000 $\AA$ break ($c_{S,0.4}$) is determined with the best accuracy ($\sim 3\%$) among all the SED basis components. This is because the 4000 $\AA$ break provides a strong spectral feature to help break the redshift-spectral degeneracy in the data (see further discussion in Sec.~\ref{S:discussion_Pk_constraints}). The power spectrum $P(k)$ can also be reconstructed unbiasedly, although we note that the posterior constraints only give moderate improvement from the prior (see further discussion on the $P(k)$ constraints in Sec.~\ref{S:discussion_Pk_constraints}). The MCMC results on $S^i$, $M^i$, and $P$ parameters do not show strong normalization degeneracy, since we have included the regularization term (Eq.~\ref{E:prior_reg}) to our likelihood function. Our algorithm also recovers the matrix $\mathbf{N}_{\ell}$ with a few percent error at low $\ell$ and the uncertainty decreases with $\ell$, since there are more independent modes on small scales. 

\begin{figure}[ht!]
\begin{center}
\includegraphics[width=\linewidth]{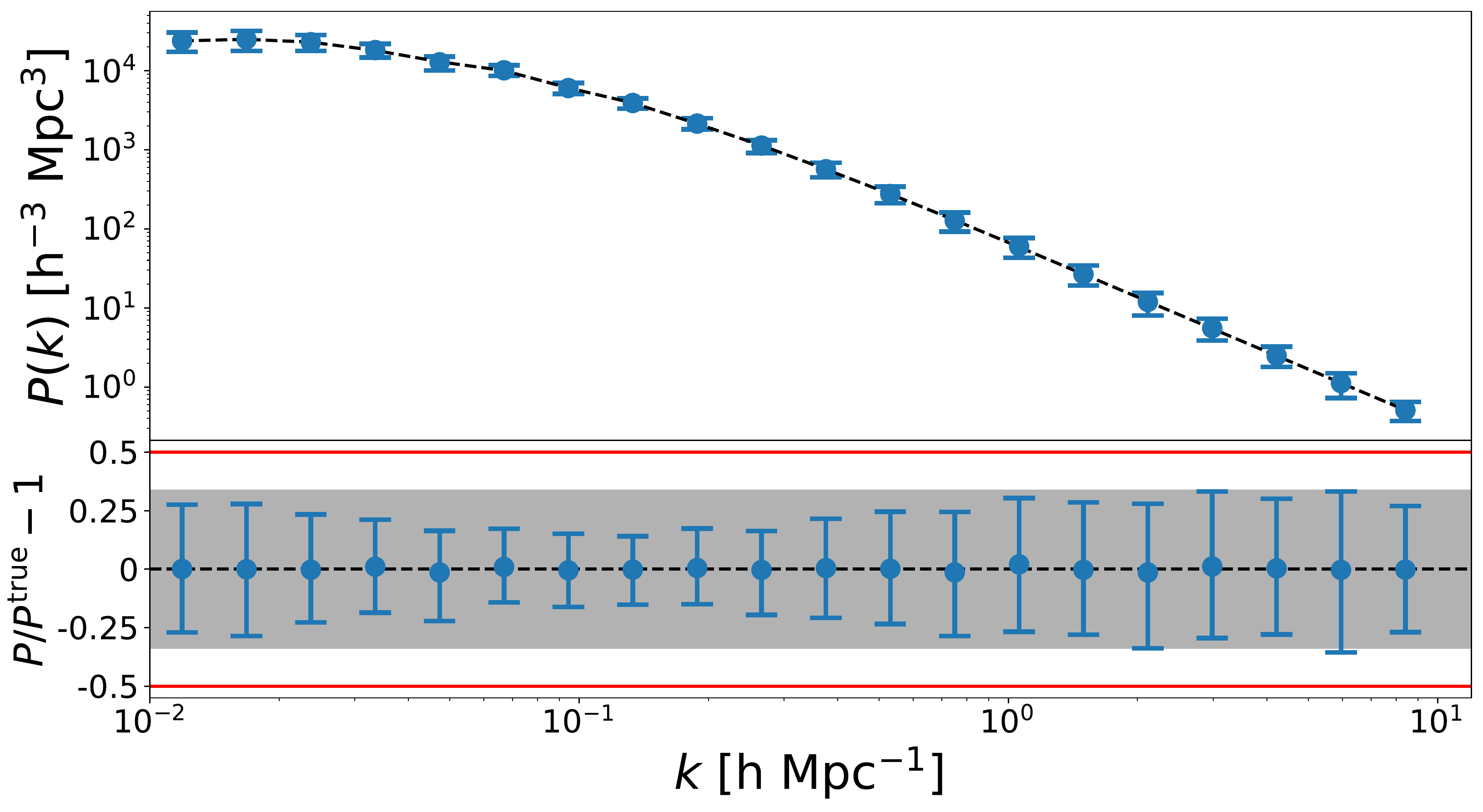}
\caption{\label{F:cv} Top: MCMC posterior on $P(k)$ with the case of the fiducial model but no sample variance fluctuations in the input data. Error bars denote the 68th percentile of the marginalized posterior, and the black-dashed line denotes the truth input model. Bottom: fractional error against the truth values. For reference, the red lines mark the prior limit of the power spectrum ($\pm50\%$ of the truth) where the prior probability is set to zero outside this range, and the gray band denotes the 68th percentile of the prior probability distribution.}
\end{center}
\end{figure}

As a sanity check, we also run another case using the same fiducial model but without sample variance fluctuations. The results of power spectrum constraints are shown in Fig.~\ref{F:cv}. Without sample variance, our MCMC posterior can recover the truth values. Also, from Fig.~\ref{F:cv}, we can see that the power spectrum has the best constraints at intermediate $k$ scales ($k\sim 0.1$ $h$ Mpc$^{-1}$), as the clustering power is suppressed on smaller scales, and the large-scale signal is susceptible to sample variance.  

\begin{figure*}[ht!]
\begin{center}
\includegraphics[width=\linewidth]{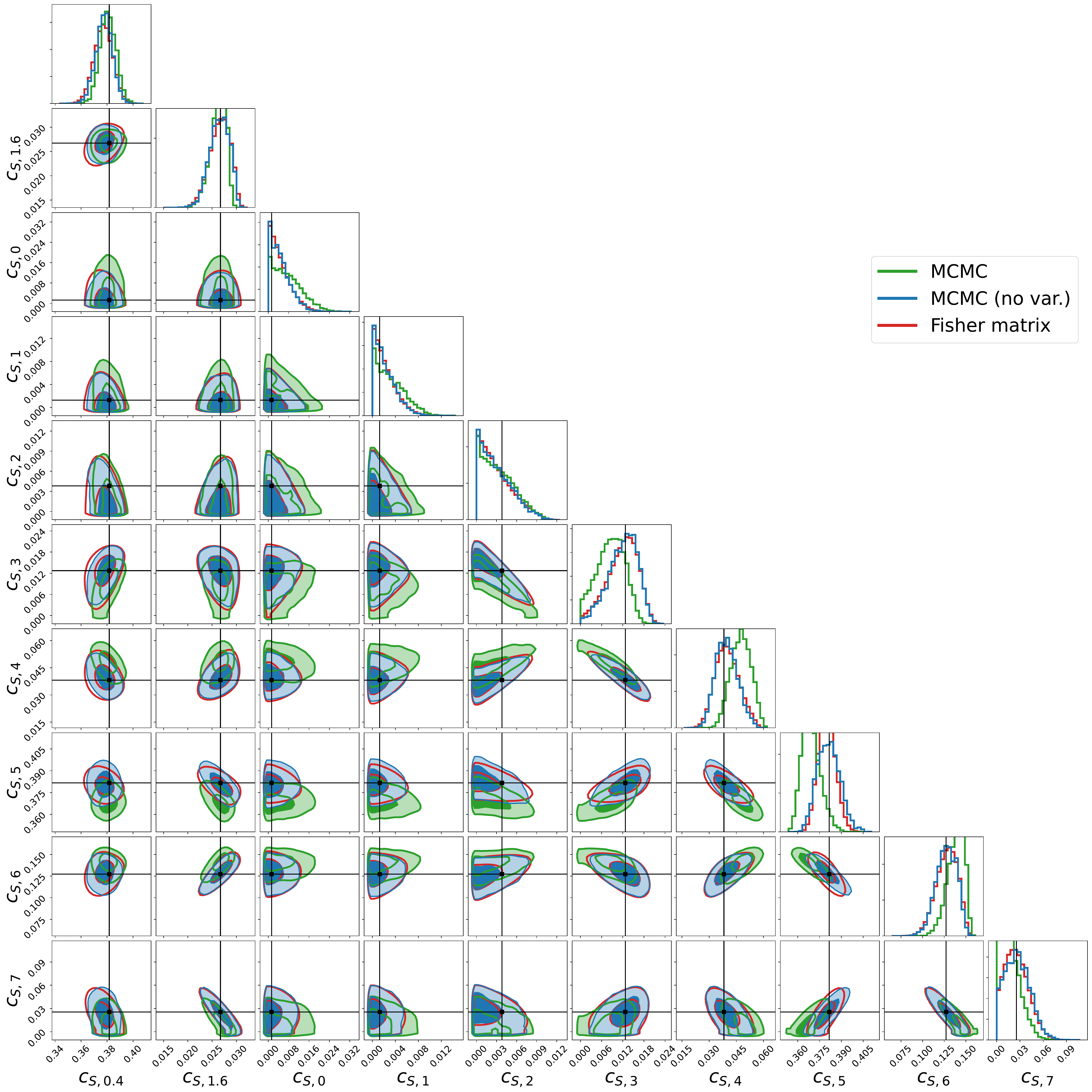}
\caption{\label{F:MCMC_S} Marginalized MCMC 2D posterior of the SED basis coefficients $c_{S,m}$ with the fiducial model (green). The blue and red contours denote the posterior of the same case but without sample variance fluctuations in the input data inferred from MCMC and the Fisher matrix, respectively. Black crosses mark the truth values. Some 2D contours have blunt edges since their posterior distributions are close to the prior limit. We apply the same boundary constraints to the Fisher matrix posterior, so it has the same blunt edges as the MCMC cases.}
\end{center}
\end{figure*}

\begin{figure}[ht!]
\begin{center}
\includegraphics[width=\linewidth]{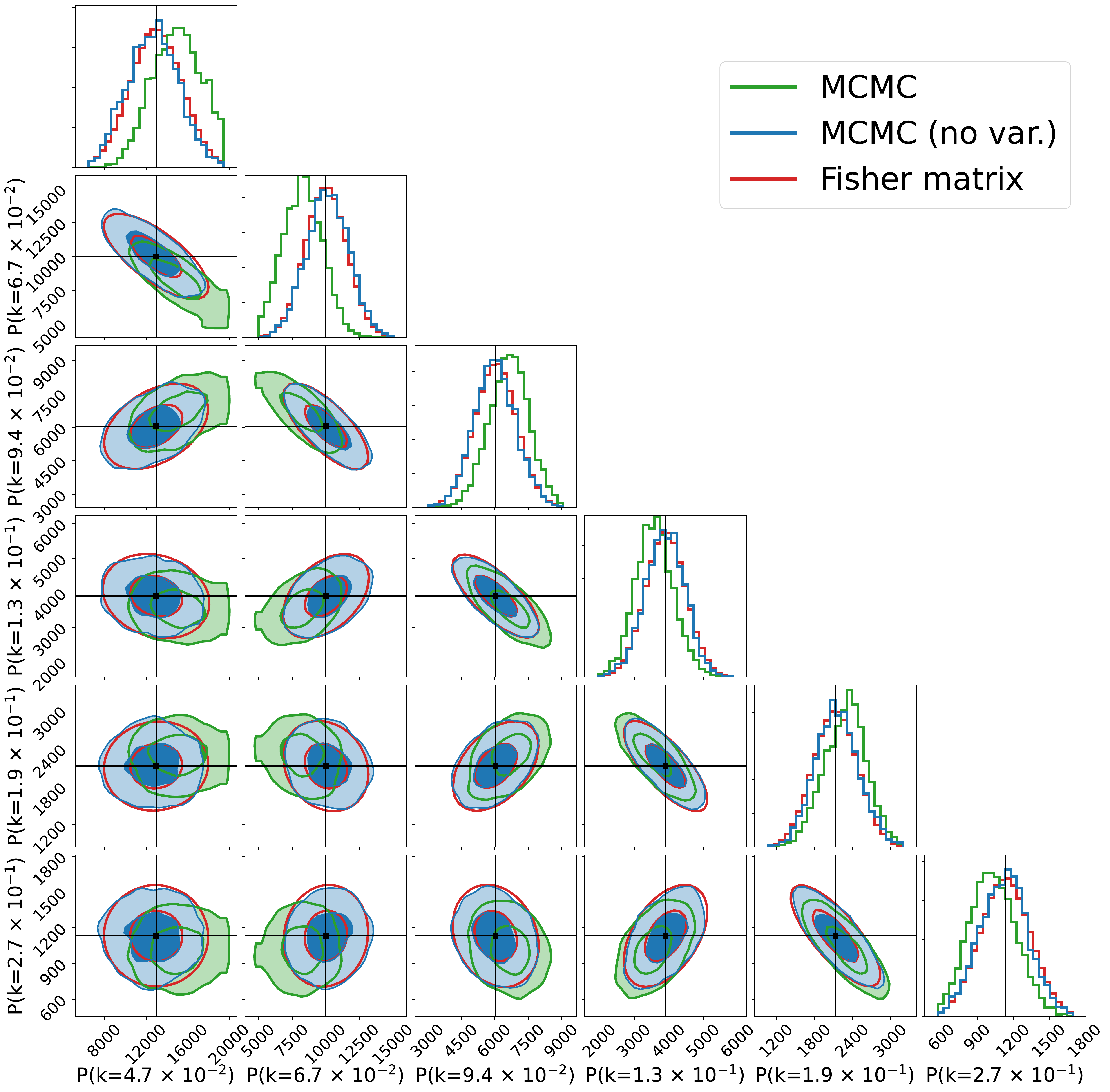}
\caption{\label{F:MCMC_P} Marginalized MCMC 2D posterior of six of the $P(k)$ modes at $k\sim 0.1$ $h$ Mpc$^{-1}$, where our model has the best power spectrum constraints. The MCMC results with the fiducial case are shown with the green contour, and the blue and red contours show the posterior of the same case but without sample variance fluctuations in the input data inferred from MCMC and the Fisher matrix, respectively. Black crosses mark the truth values. Some 2D contours have blunt edges since their posterior distributions are close to the prior limit.}
\end{center}
\end{figure}

Fig.~\ref{F:MCMC_S} and Fig.~\ref{F:MCMC_P} show the 2D posterior of SED coefficients $c_S$ and six of the $P(k)$ modes near $k=0.1$ $h$ Mpc$^{-1}$, where the posterior has the best parameter constraints. Our MCMC results of the case without sample variance fluctuations are consistent with the truth values, and the covariance is in agreement with the analytic expression from the Fisher matrix. From the posterior distribution, we find almost no covariance between noise and the parameters in $S^i$, $M^i$, and $P$, and only a very small covariance between $P(k)$ and the $S^i$ and $M^i$ coefficients, whereas there is non-negligible covariance between $S^i$ and $M^i$ coefficients. This can be understood by the form of the angular power spectrum in Eq.~\ref{E:Cl}: the noise $\mathbf{N}_\ell$ is a separate additive term to the clustering signal, and thus it has small correlations with the $S^i$, $M^i$, and $P$ terms; the $S^i$ and $M^i$ are highly mixed in $\mathbf{B}_\ell$ through the line-of-sight integration (Eq.~\ref{E:Bl}), so they are expected to be strongly correlated.

\section{Discussion}\label{S:discussion}
\subsection{Information from Different $\ell$ Modes}\label{S:discussion_fisher}
\begin{figure}[ht!]
\begin{center}
\includegraphics[width=\linewidth]{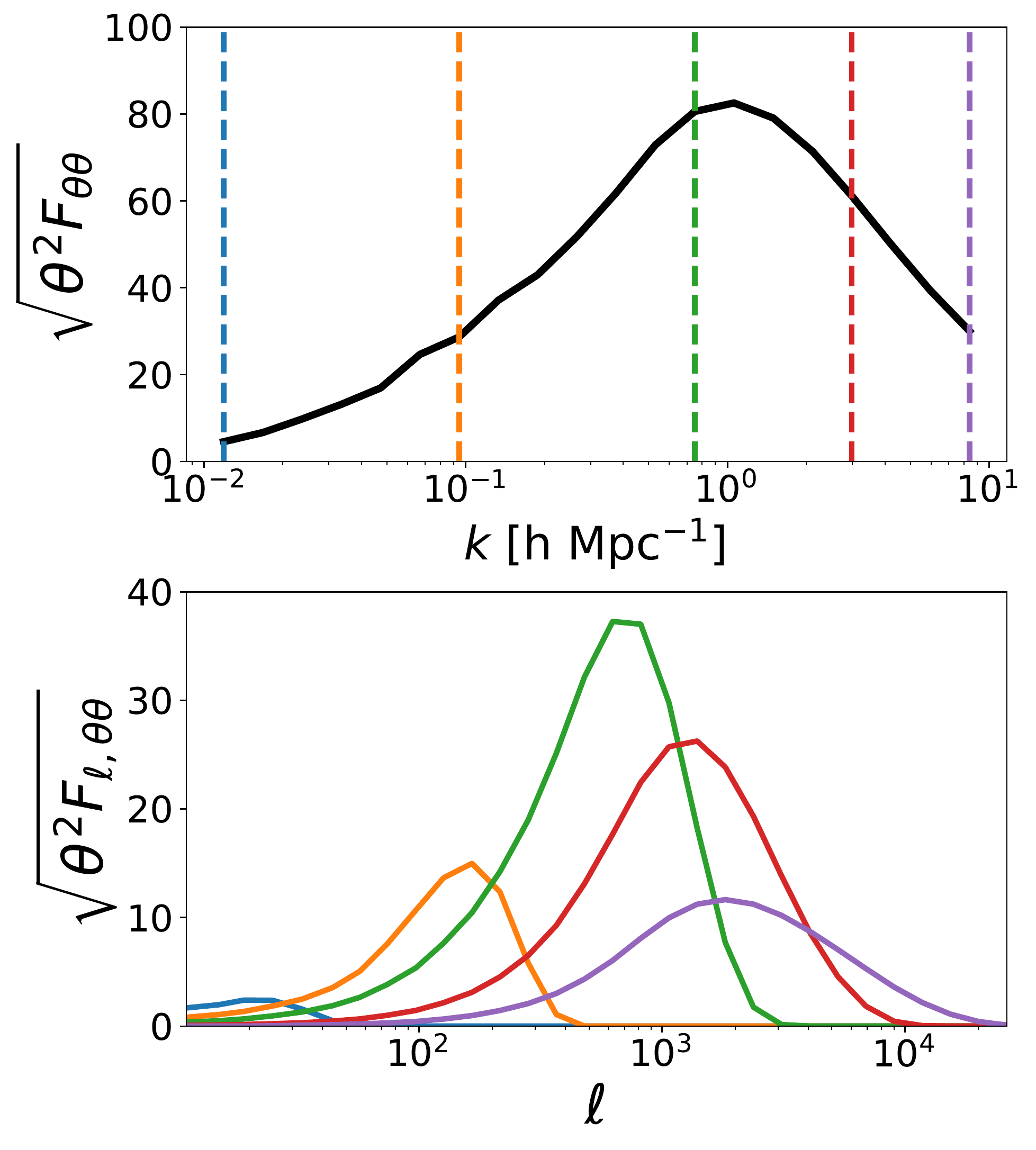}
\caption{\label{F:discussion_fisher} Top: diagonal elements of Fisher matrix on power spectrum $P(k)$ parameters with the fiducial case. The Fisher matrix element $F_{\theta\theta}$ is the inverse variance of parameter $\theta$ given other parameters fixed at the truth values, and thus $\sqrt{\theta^2 F_{\theta\theta}}$ represents the signal-to-noise ratio on $\theta$. Bottom: Fisher information from individual $\ell$ modes on parameter $\theta$, $F_{\ell,\theta\theta}$. We show five of the power spectrum modes where their $k$ values are marked as vertical dashed lines in the top panel.}
\end{center}
\end{figure}
The Fisher information matrix measures the information content of the data. As each $\ell$ mode is independent, the total Fisher information $\mathbf{F}$ is given by the sum of the Fisher information from individual $\ell$ modes $\mathbf{F}_\ell$ (Eq.~\ref{E:Fisher}), so $\mathbf{F}_\ell$ can be used to quantify information content from data $\mathbf{C}_\ell^d$ at mode $\ell$. However, $\mathbf{F}_\ell$ are singular matrices in our formalism, so they cannot be directly inverted to obtain the covariance matrices. Therefore, we instead evaluate the constraint on parameter $\theta$ with its diagonal Fisher matrix element, $\mathbf{F}_{\ell,\theta\theta}$. Note that the diagonal elements of the Fisher matrix are the inverse variance on parameter $\theta$ given other parameters fixed at the truth value instead of marginalized over other parameters. An example with the fiducial model is shown in Fig.~\ref{F:discussion_fisher}. The top panel shows the total Fisher information on $P(k)$ from all $\ell$ modes, and we show $\sqrt{\theta^2 F_{\theta\theta}}$ to represent the signal-to-noise ratio on each parameter given other parameters fixed at the truth value. The bottom panel breaks down the information from different $\ell$ modes in five selected $k$ bins, and we can see the correspondence of angular modes $\ell$ and Fourier modes in co-moving space $k$ with our fiducial setup that considers emission from $0<z<3$.

\subsection{Power Spectrum Constraints}\label{S:discussion_Pk_constraints}
Here, we discuss how the constraints on the 3D power spectrum $P(k)$ depend on different factors. In our data, the power spectrum $P(k)$ is projected to 2D spectral-intensity maps with the projection kernel at each frequency determined by $S$ and $M$. Therefore, the $S$, $M$, and $P$ signals are highly confused in the data, which means any information that breaks this confusion will greatly improve the $P(k)$ constraints.

\begin{figure}[ht!]
\begin{center}
\includegraphics[width=\linewidth]{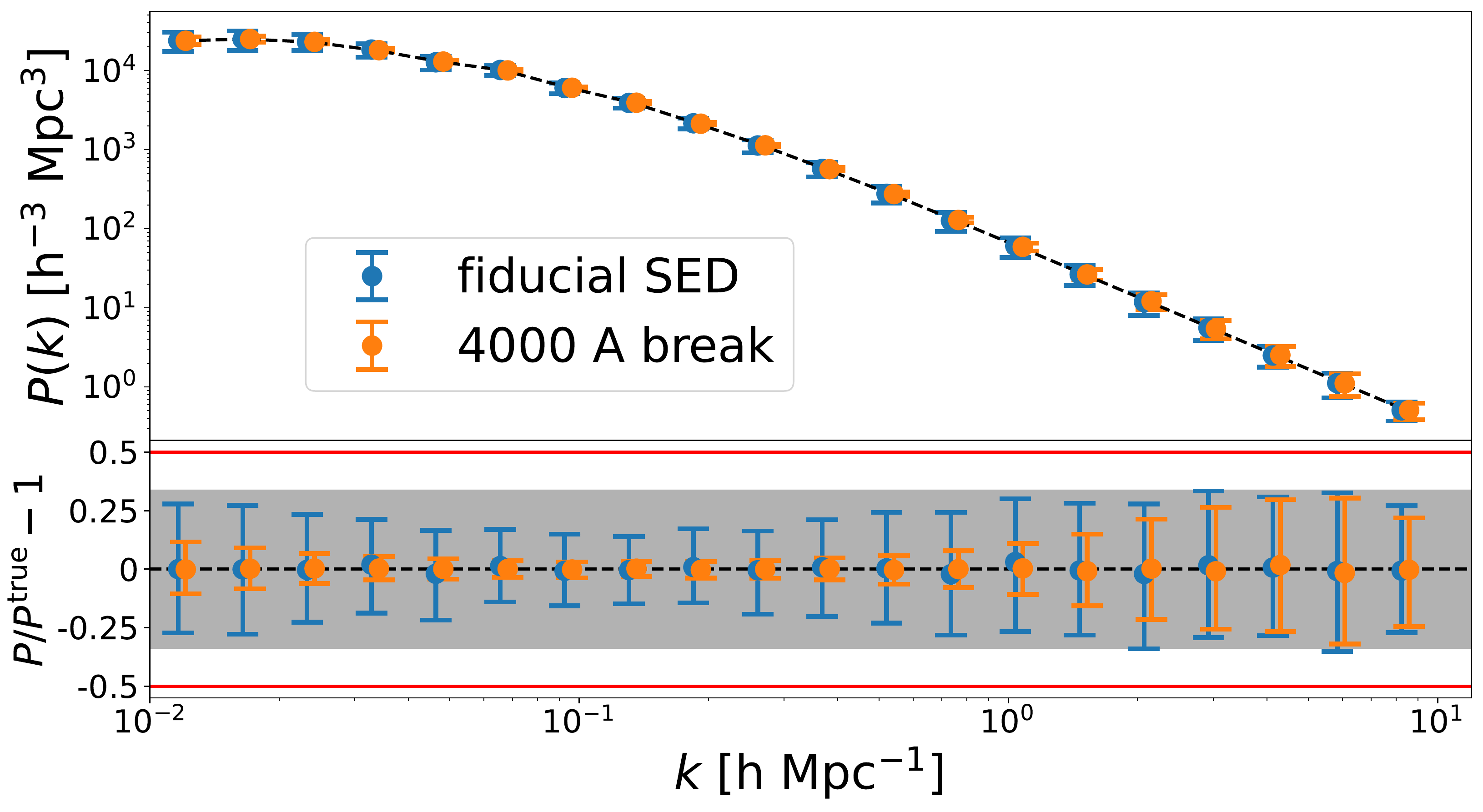}
\caption{\label{F:discussion_4000_P} Top: MCMC posterior on $P(k)$ with the case of a pure 4000 $\AA$ break SED (orange) compared to the fiducial case (blue). For better visualization, the input data for both cases does not include the sample variance fluctuations. Error bars denote the 68th percentile of the marginalized posterior, and the black-dashed line denotes the truth input model. Bottom: fractional error against the truth values. For reference, the red lines mark the prior limit of the power spectrum ($\pm50\%$ of the truth) where the prior probability is set to zero outside this range, and the gray band denotes the 68th percentile of the prior probability distribution.}
\end{center}
\end{figure}

From Fig.~\ref{F:S}, we can see that the 4000 $\AA$ break is the best-constrained component in the SED in our fiducial case, as the spectral break feature allows for unambiguously determining the redshift of emission sources to reconstruct the three-dimensional distribution traced by them. To further demonstrate this effect, we run a case by replacing the SED of the fiducial model with a pure 4000 $\AA$ break spectrum, i.e., the Heaviside function with the step at 4000 $\AA$ (Eq.~\ref{E:Sbasis_4000}). The results of the power spectrum constraints are shown in Fig.~\ref{F:discussion_4000_P}. For better visualization, we show the case without sample variance fluctuations. With the pure 4000 $\AA$ SED, the posterior constraints are significantly better than the fiducial case.

\begin{figure}[ht!]
\begin{center}
\includegraphics[width=\linewidth]{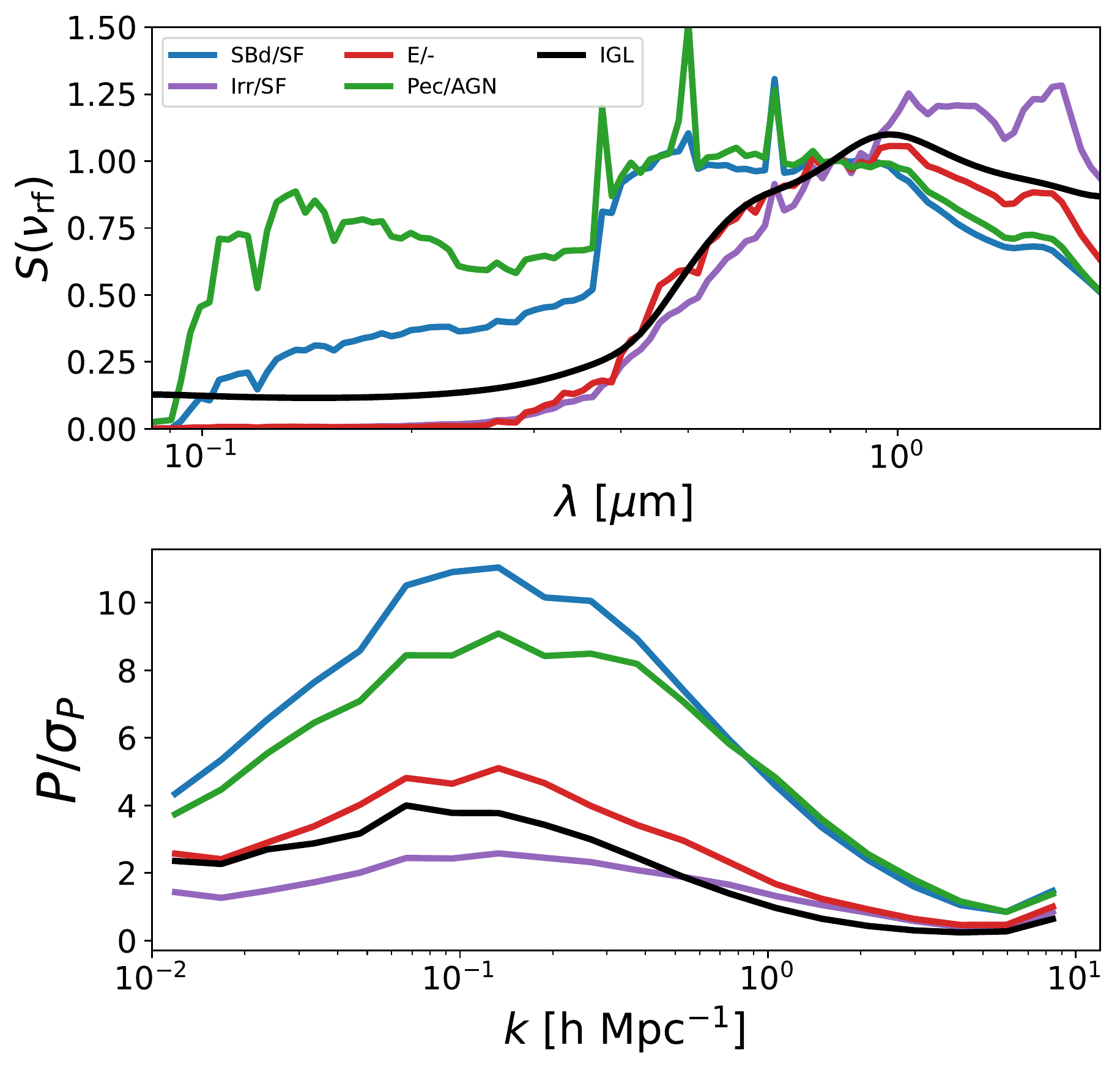}
\caption{\label{F:discussion_obs}Top: the SEDs of the H12 IGL model and four local galaxies. See Sec.~\ref{S:signal_model} and Fig.~\ref{F:model_S} for details. Bottom: the signal-to-noise ratio of the power spectrum $P(k)$ of the five SED cases shown in the top panel. Here, all parameters are set to the fiducial case except for $S^i(\nu_{\rm rf})$, where we use 100 logarithmically spaced frequency bins spanning $0.33$--$2$ $\mu$m. The signal-to-noise ratio is calculated with the Fisher matrix.}
\end{center}
\end{figure}

We further investigate the power spectrum constraints with different source SEDs. We consider five SEDs discussed in Sec.~\ref{S:signal_model}, where one of them is the H12 model of IGL from $z=0$ sources, and the other four are the SEDs from local galaxies with different galaxy types \citep{2014ApJS..212...18B}. Instead of using the set of 10 basis functions as in our fiducial case, we use 100 logarithmically spaced frequency bins spanning $0.33$--$2$ $\mu$m, which is the full rest-frame spectral range that will be probed by the nine photometric bands we considered from redshift $0<z<3$. Using the SED values in 100 spectral bins as our $S^i(\nu_{\rm rf})$ parameters allows us to capture fine features in these SEDs such as spectral lines. Other components ($M$, $P$, and $\mathbf{N}_\ell$) are set to the fiducial case in this test. Fig.~\ref{F:discussion_obs} shows the signal-to-noise ratio of the binned 3D power spectrum $P(k)$ from the Fisher matrix. The variance $\sigma_P^2$ is the diagonal elements of the inverse Fisher matrix. The SEDs of the SBd/SF (blue) and Pec/AGN (green) cases give much tighter constraints on $P(k)$ since they have stronger sharp features (4000  $\AA$ break and spectral lines) that help distinguish the redshift of emitting sources. 

In addition, comparing the E/- (red) and Irr/SF (purple) cases in Fig.~\ref{F:discussion_obs}, we find weaker large-scale (low-$k$) $P(k)$ constraints for the Irr/SF case. This is because this case has a much stronger emission on the long-wavelength rest-frame SED, which results in more weighting toward low-redshift emission in the data, whereas the large-scale $P(k)$ are more sensitive to the high-redshift signal. This also indicates that depending on the scale of interests and the SEDs of the sources, there will be an optimal set of observing filters to better constrain the power spectrum.

\begin{figure}[ht!]
\begin{center}
\includegraphics[width=\linewidth]{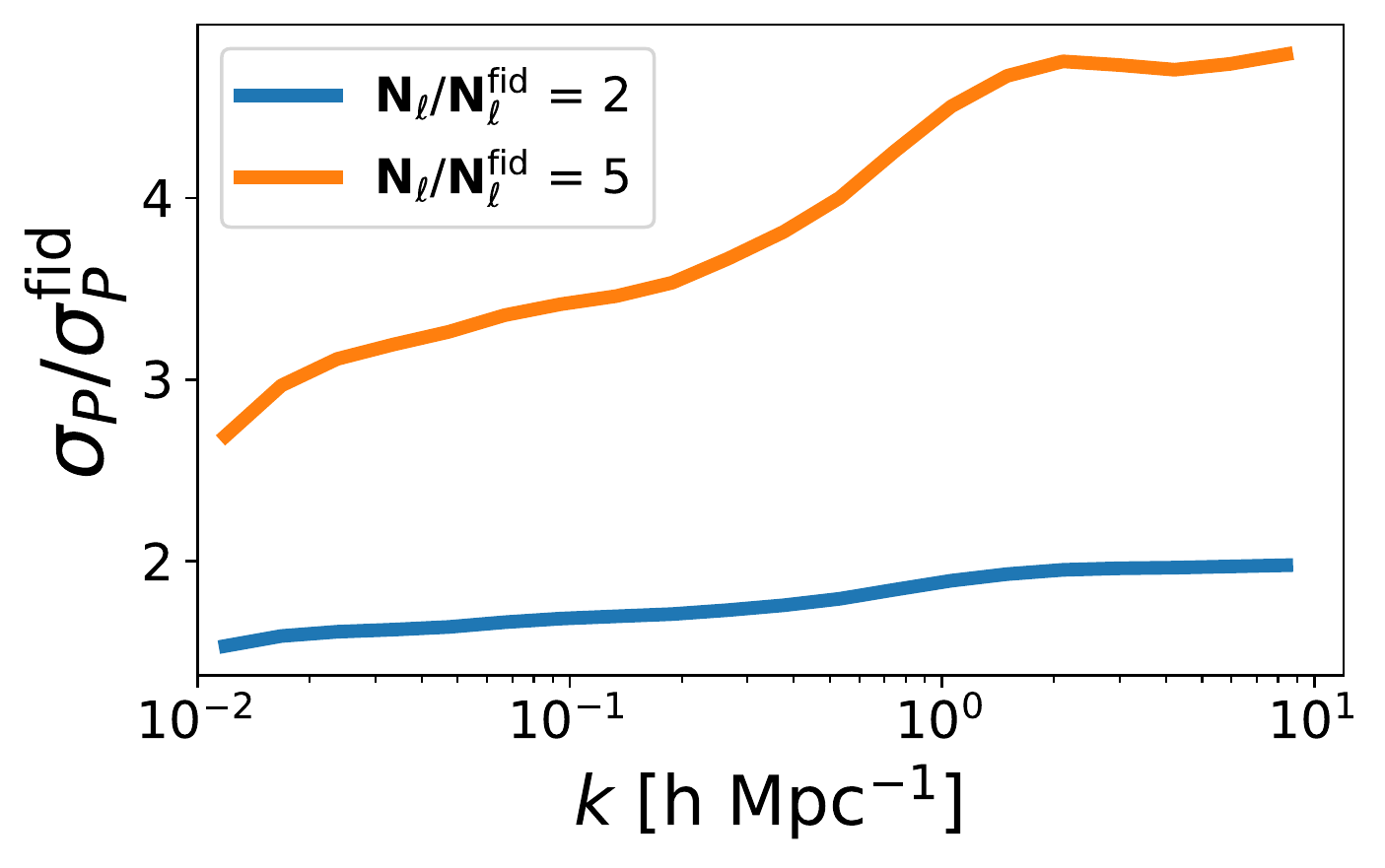}
\caption{\label{F:Fisher_Nscale} Ratio of the $P(k)$ uncertainties of the case with increased noise level compared to the fiducial case. In the blue/orange case, we set the noise power spectrum $\mathbf{N}_\ell$ to $2/5$ times higher than the fiducial case in all frequency bands. The uncertainties are calculated with the Fisher matrix.}
\end{center}
\end{figure}

Finally, we also investigate the dependence of power spectrum constraints on the noise. Fig.~\ref{F:Fisher_Nscale} shows the $P(k)$ uncertainties compared to the fiducial case when we increase the noise power spectrum $\mathbf{N}_\ell$ by a factor of $2$ and $5$, respectively. The noise affects the $P(k)$ constraints significantly on all scales, while the small-scale (high-$k$) modes are more sensitive to the noise due to their smaller clustering-to-noise ratio in the power spectrum. We note that in reality, the Poisson noise also depends on $S$ and $M$, and it has cross-frequency correlations, so the $P(k)$ dependence on the noise level might be different from the case considered here. Further investigation with the full Poisson noise prescription will be studied in future work.

\begin{figure}[hb!]
\begin{center}
\includegraphics[width=\linewidth]{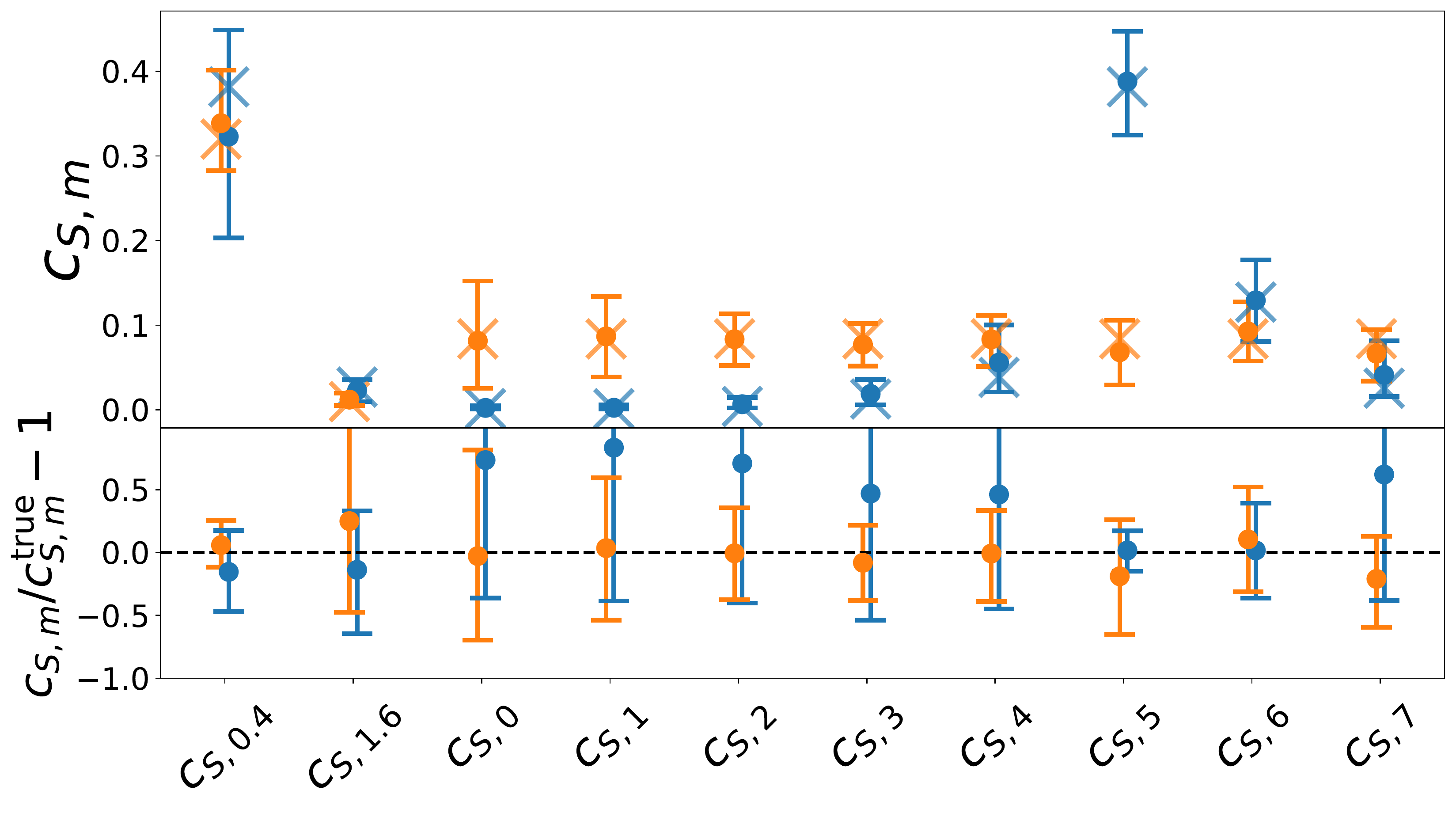}
\includegraphics[width=\linewidth]{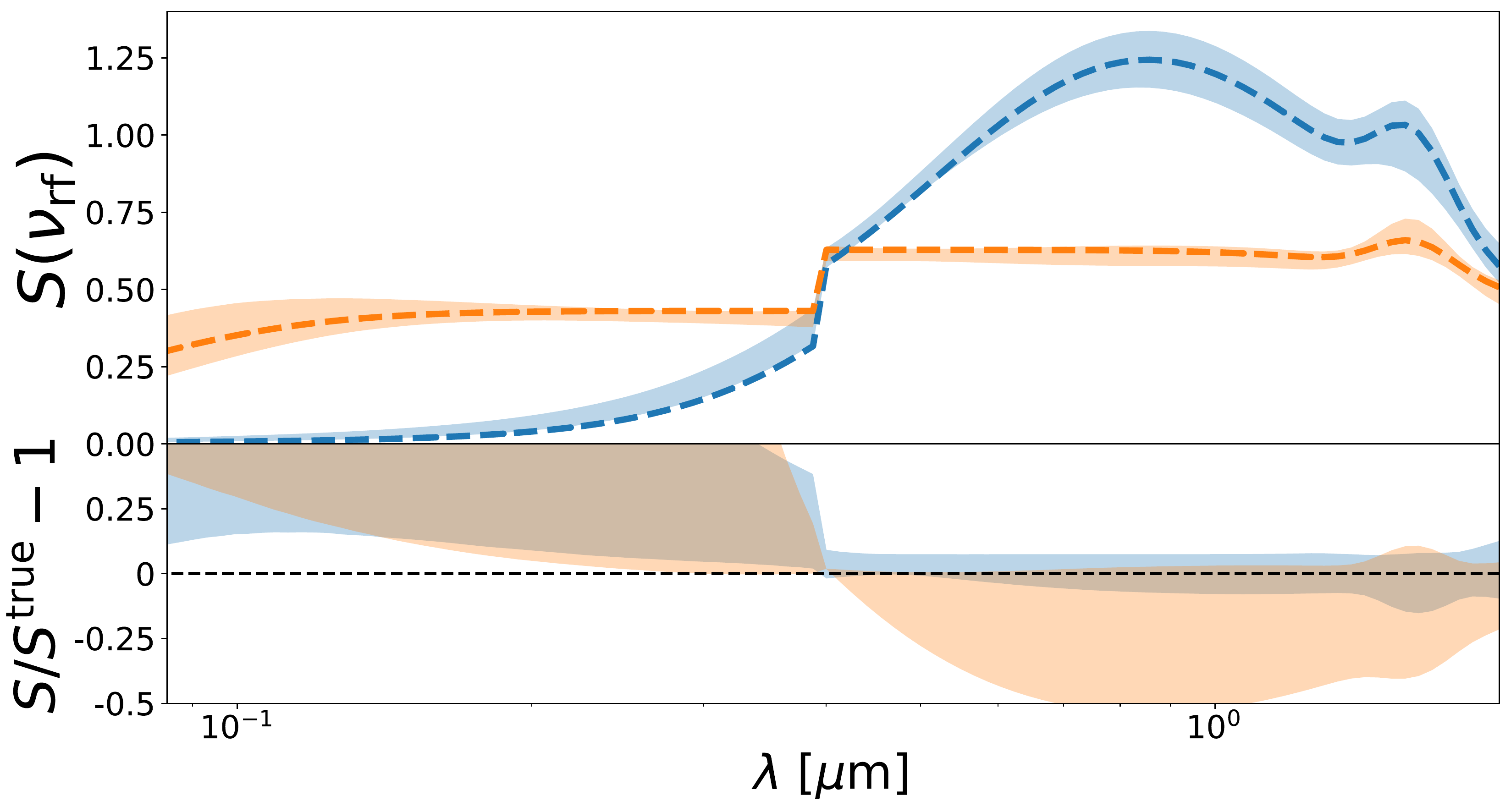}
\caption{\label{F:discussion_2cls_S} MCMC posterior on $S^i(\nu_{\rm rf})$ with the two components case without sample variance fluctuations. All parameters for the first component and the survey setup are identical to the fiducial case. Top: constraints on the $S^i(\nu_{\rm rf})$ basis function amplitudes, $c_{S,m}$, for the first (blue) and the second (orange) components. Error bars denote the 68th percentile of the marginalized posterior, and the crosses denote the truth values. Bottom: the 68th posterior percentile of the reconstructed source SED $S^i(\nu_{\rm rf})$ for the first (blue-shaded region) and second (orange-shaded region) components. The blue/orange dashed lines denote the truth input mode of the first/second components. The bottom subpanel in each plot show the fractional error against the truth values.}
\end{center}
\end{figure}

\subsection{Multiple Source Components}\label{S:discussion_2cls}
\begin{figure}[ht!]
\begin{center}
\includegraphics[width=\linewidth]{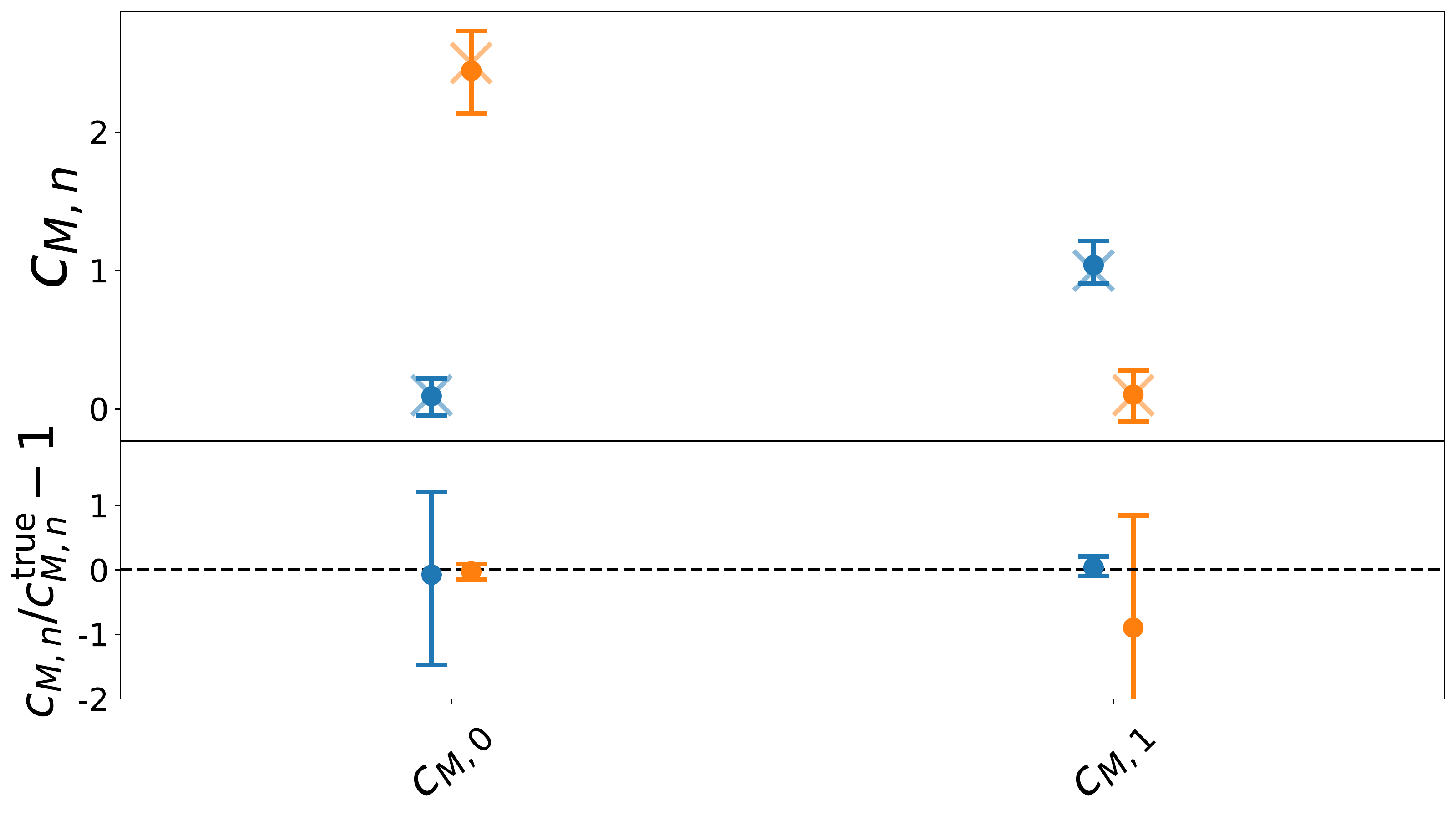}
\includegraphics[width=\linewidth]{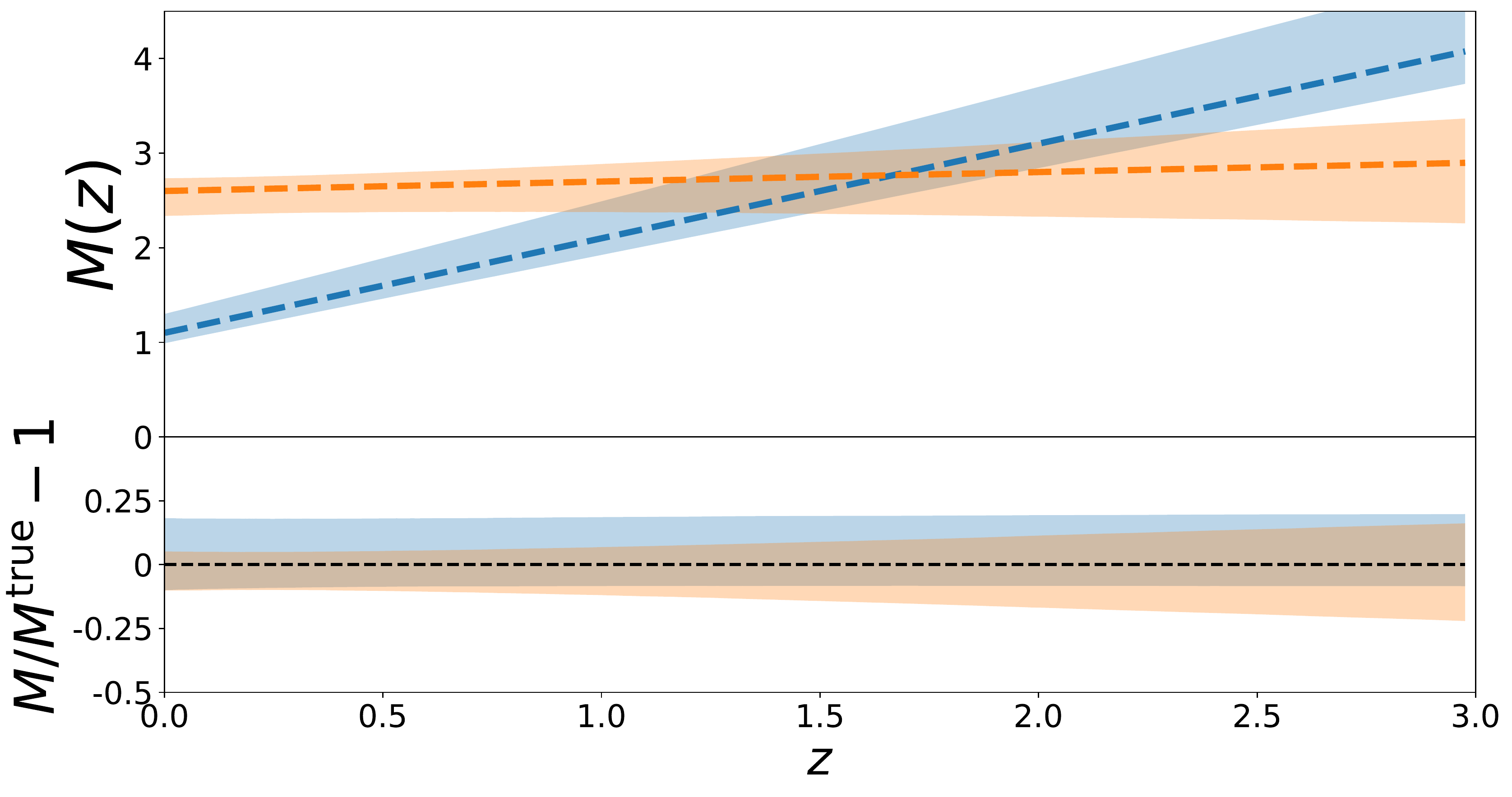}
\caption{\label{F:discussion_2cls_M} MCMC posterior on $M^i(\chi)$ with the two-component case without sample variance fluctuations. All parameters for the first component and the survey setup are identical to the fiducial case. Top: constraints on the $M^i(\chi)$ basis function amplitudes, $c_{M,n}$, for the first (blue) and the second (orange) components. Error bars denote the 68th percentile of the marginalized posterior, and the crosses denote the truth values. Bottom: the 68th posterior percentile of the reconstructed $M^i(\chi)$ for the first (blue-shaded region) and second (orange-shaded region) components. The blue/orange dashed lines denote the truth input mode of the first/second components. The bottom subpanel of each plot shows the fractional error against the truth values.}
\end{center}
\end{figure}

\begin{figure}[ht!]
\begin{center}
\includegraphics[width=\linewidth]{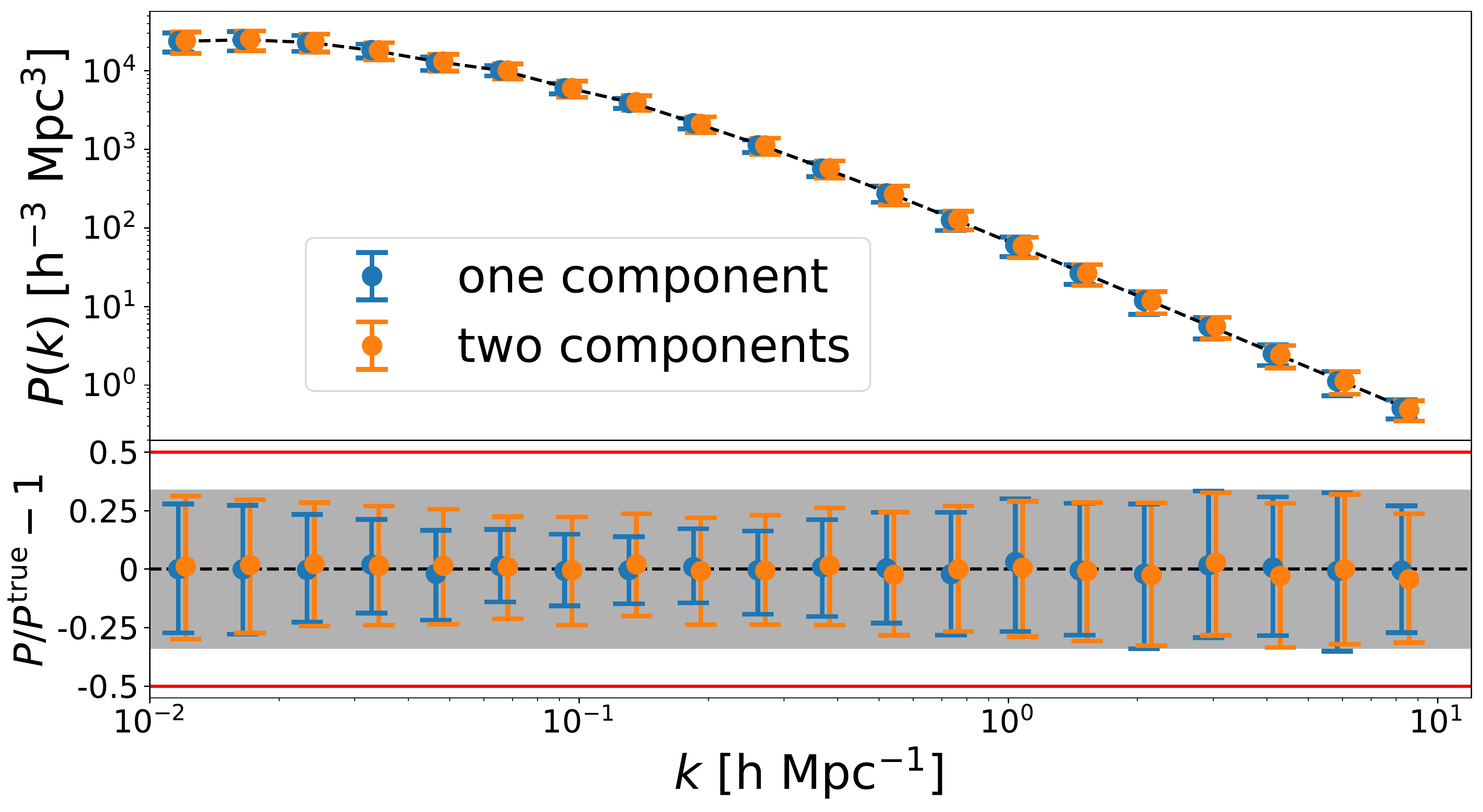}
\caption{\label{F:discussion_2cls_P} Top: MCMC posterior on $P(k)$ with the case of two components (orange) compared to the fiducial single-component case (blue). For better visualization, the input data for both cases does not include the sample variance fluctuations. Error bars denote the 68th percentile of the marginalized posterior, and the black-dashed line denote the truth input model. Bottom: fractional error against the truth values. For reference, the red lines mark the prior limit of the power spectrum ($\pm50\%$ of the truth) where the prior probability is set to zero outside this range, and the gray band denotes the 68th percentile of the prior probability distribution.}
\end{center}
\end{figure}

\begin{figure}[ht!]
\begin{center}
\includegraphics[width=\linewidth]{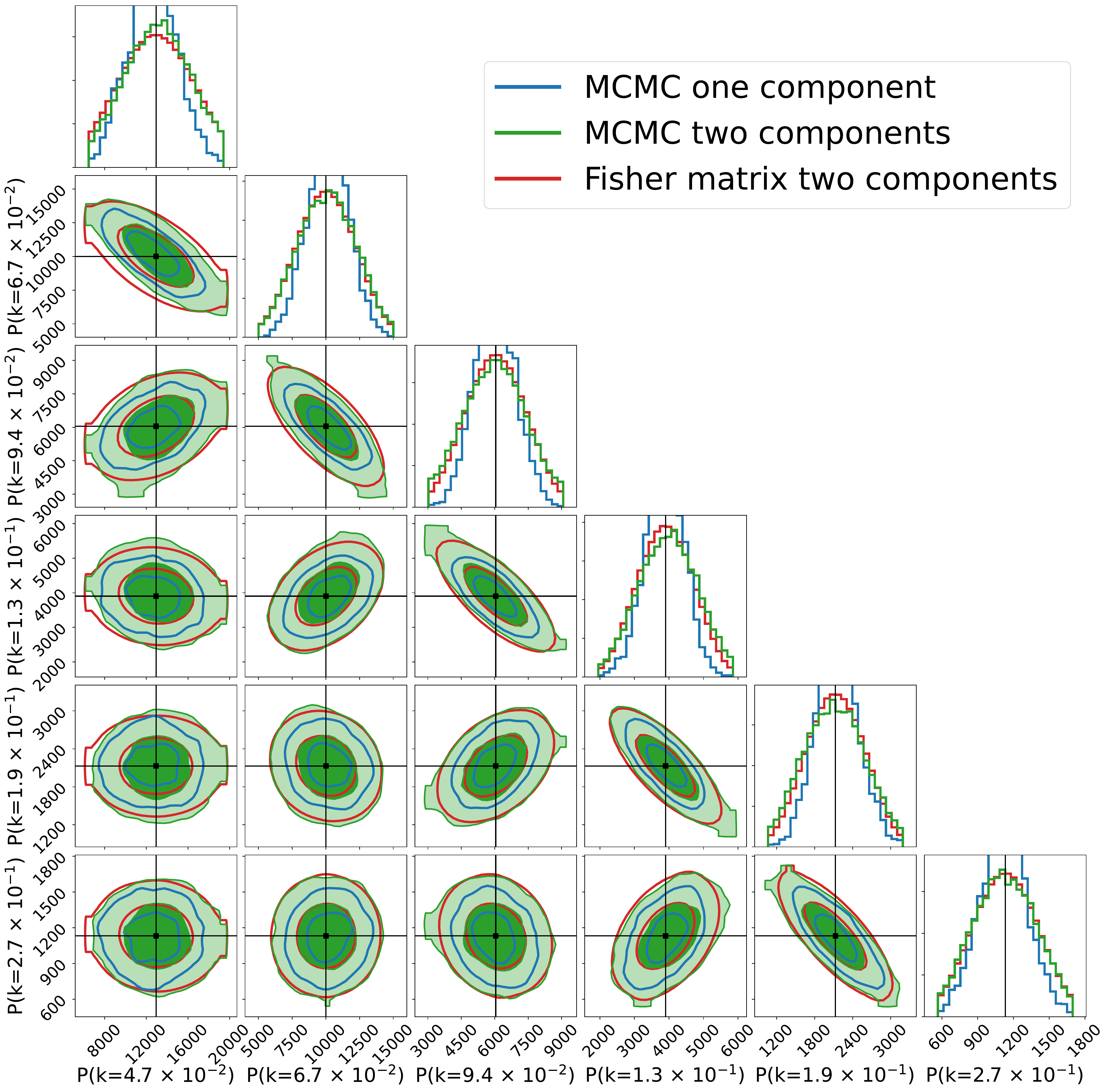}
\caption{\label{F:discussion_2cls_MCMC_P} Marginalized MCMC 2D posterior of the two-component case without sample variance fluctuations (green). Here, we show six of the $P(k)$ modes at $k\sim 0.1$ $h$ Mpc$^{-1}$, where our model has the best power spectrum constraints. The red contours mark the posterior from the Fisher matrix of the same case. For comparison, the blue contours show the constraints from the one-component case (i.e. the blue contours in Fig.~\ref{F:MCMC_P}). Black crosses mark the truth values. Some 2D contours have blunt edges since their posterior distributions are close to the prior limit. We apply the same boundary constraints to the Fisher matrix posterior, so it has the same blunt edges as the MCMC cases.}
\end{center}
\end{figure}

Here, we present a case of two components ($N_c=2$), where we use the same fiducial model for the first component, and add another signal component with a smoother SED and luminosity density functions. In this case, we have $N_\theta=314$ parameters. The sample variance fluctuations are not included here to better compare the constraints with the fiducial one-component case. The MCMC results on $S^i$, $M^i$, and $P$ are shown in Fig.~\ref{F:discussion_2cls_S}, \ref{F:discussion_2cls_M}, and \ref{F:discussion_2cls_P}, respectively. With the additional degree of freedom from multiple source components, the constraints on all $S^i$, $M^i$, and $P$ parameters are degraded compared to the fiducial one-component case. The 2D posterior of a few selected $P(k)$ at the modes near $k=0.1$ $h$ Mpc$^{-1}$ is shown in Fig.~\ref{F:discussion_2cls_MCMC_P}. Our MCMC results are consistent with the analytic expression from the Fisher matrix. We also verify that our results from the same case with sample variance fluctuations in the data give unbiased parameter constraints.

\subsection{Number of Components}\label{S:number_of_components}
In Sec.~\ref{S:intensity_field}, we describe our formalism by classifying individual sources into different components. In reality, our algorithm finds a set of SED components that best describes the aggregate emission field without the notion that the underlying signal is emitted by discrete sources. Therefore, the resulting SEDs will not necessarily correspond to any SED of individual sources; instead, our process will pick out dominant features from the set of SEDs as our components. This is similar to the concept of principal component analysis (PCA), which summarizes data with principal modes.

To assess the number of components $N_c$ required in reality, we perform a PCA on an SED library \citep{2009ApJ...690.1236I}, and found that the variety of SEDs can be well captured by about 10 to 20 PCA modes. Although our formalism is not equivalent to this test, we expect the same order of magnitude (a few tens) of $N_c$ is needed for a realistic survey. We leave a more detailed investigation to future papers.

\section{Unique Advantages of Our Method}\label{S:discussion_advantages}
\subsection{Flexibility}
While we need to parametrize our signal in the inference, our framework is flexible to use with any parametrization scheme without any prior assumption on the signal or noise. This assumption-free analysis framework can avoid biases from insufficient modeling, which is an inherent issue for many existing methods. For example, photometric redshift surveys will rely on a known set of SED templates to infer the redshift of their sources. However, the high-redshift galaxy SED might not be consistent with any SED in a template bank built from lower-redshift samples. Our approach has the flexibility to discover signals that are not in current models to overcome the modeling bias, as well as to utilize information from those sources. This is crucial for future surveys, as they are expected to achieve higher sensitivity to probe the faint and distant populations over a wide range in redshift.

In addition, any prior information can also be included in our analysis. For example, if we know the SED for some sources in the data, we can fix one of the $S^i$ components to that SED to reduce the number of free parameters in the fitting. Similarly, for the power spectrum $P(k)$, instead of fitting the power on discrete $k$ bins, we can parametrize it with a combination of a few smooth functions to restrict the smoothness of $P(k)$. Finally, the correlation between parameters can also be specified by including the parameter covariance in the prior function.

\subsection{Generalizability}
In this work, we only demonstrate our method with spectral-intensity maps, but we emphasize that this is a general framework that can be applied to any other LSS tracer (or its combinations). For example, we can combine the spectral-intensity maps with a 3D galaxy catalog generated from the same data set or from any other surveys observing the same sky region with arbitrary depth. This can be done by formulating their auto and cross power spectra and their likelihood function on parameters, and derive the joint constraints from these two datasets. We leave this analysis to future work.

\section{Conclusion and Future Work}\label{S:conclusion}
We present a novel technique to analyze large-scale cosmological survey data. In contrast to conventional detection-based galaxy surveys, our method infers underlying large-scale structures, properties of emission sources, and the noise, directly from spectral-intensity maps without resolving individual sources. We use a data-driven approach to constrain the signal solely from data covariance (i.e., auto and cross angular power spectrum $C_{\ell,\nu\nu'}$) without any external information, and we only use the assumptions of the signal homogeneity and isotropy and the fact that a finite number of source components can fully capture the emission field. This method allows us to fully exploit information that lies in the data when the emission field is Gaussian and can be fully characterized by two-point statistics, which is true for the large-scale cosmological signal.

As a proof of principle, we consider an observation from nine photometric bands in the optical and near-infrared, and the emissions from a single component of sources. We show that our algorithm can reconstruct the input source SED, luminosity density, underlying 3D power spectrum $P(k)$, and noise from all combinations of auto and cross angular power spectrum $C_{\ell,\nu\nu'}$. We also present a case with two components of sources, and demonstrate that our algorithm can infer the input model and characterize parameter uncertainties in this case too.

We quantify uncertainties on parameter constraints using a Bayesian framework using both MCMC and a semi-analytical approach based on the Fisher information matrix, and verified that the two methods give consistent results.

We investigate the information on the 3D power spectrum $P(k)$ from different angular modes, and find a strong correspondence of angular ($\ell$) and spatial ($k$) scales. We explore cases with different source SED, and find that SEDs with sharp features (spectral breaks or emission lines) give strong constraining power on the 3D power spectrum $P(k)$, as these features help anchor the redshift of emitting sources. 

This paper focuses on establishing the formalism for modeling signals in a light cone, and showcasing the potential of extracting LSS signals directly from spectral-intensity maps in a data-driven manner. While the models and example cases presented in this work are simple scenarios, we plan to explore this data-driven approach to analyze multifrequency large-scale maps with more realistic setups in future work. This will serve as an important technique for upcoming cosmological surveys such as SPHEREx, Rubin Observatory, Euclid, or the Nancy Grace Roman Space Telescope.

\acknowledgments
Y.-T.C. acknowledges support by NASA ROSES grant 18-2ADAP18-0192. B.D.W. acknowledges support by the ANR BIG4 project, grant ANR-16-CE23-0002 of the French Agence Nationale de la Recherche; and the Labex ILP (reference ANR-10-LABX-63) part of the Idex SUPER, and received financial state aid managed by the Agence Nationale de la Recherche, as part of the programme Investissements d’avenir under the reference ANR-11- IDEX-0004-02. The Flatiron Institute is supported by the Simons Foundation. T.-C.C. acknowledges support by the JPL R\&TD initiative on line intensity mapping. Part of this work was done at Jet Propulsion Laboratory, California Institute of Technology, under a contract with the National Aeronautics and Space Administration.

\software{
astropy \citep{2013A&A...558A..33A,2018AJ....156..123A}, 
emcee \citep{2013PASP..125..306F}, corner \citep{corner}.
}

\clearpage
\appendix

\section{Power Spectrum}\label{A:power_spectrum}
\subsection{Auto and Cross angular Power spectrum}
Here, we present the derivation of the angular power spectra of the intensity field $\nu I_\nu(\nu, \hat{n}) $ (Eq.~\ref{E:nuInui_SMA}).

The mean specific intensity, $\nu I_\nu(\nu)$, in an angular area $\Omega$ is
\begin{equation}
\begin{split}
\nu I_\nu(\nu) &= \frac{1}{\Omega}\int_{\Omega}d\hat{n}\, \nu I_\nu(\nu, \hat{n})\\
&=\sum_{i=1}^{N_c}\int d\chi\, S^i(\nu_{\rm rf})M^i_0(\chi)A(\chi),
\end{split}
\end{equation}
where $M^i_0(\chi)$ is the mean luminosity of component $i$:
\begin{equation}
M^i_0(\chi) = \frac{1}{\Omega}\int_{\Omega}d\hat{n}\, M^i_0(\chi,\hat{n}).
\end{equation}
The intensity contrast is defined by
\begin{equation}
\begin{split}
\delta&\left (\nu I_\nu(\nu,\hat{n})  \right )=\nu I_\nu(\nu,\hat{n})- \nu I_\nu(\nu)\\
&=\sum_{i=1}^{N_c}\int d\chi\, S^i(\nu_{\rm rf})\delta 
M^i_0(\chi,\hat{n})A(\chi),
\end{split}
\end{equation}
where the luminosity density contrast, $\delta M^i_0(\chi,\hat{n})$, traces the underlying matter density field, $\delta_m$, on large scales with bias $b^i$:
\begin{equation}
\begin{split}
\delta M^i_0(\chi,\hat{n}) &= M^i_0(\chi,\hat{n}) - M^i_0(\chi) \\
&= b^i(\chi)M^i_0(\chi)\delta_m(\chi,\hat{n}).
\end{split}
\end{equation}
We define the bias-weighted luminosity density as
\begin{equation}
M^i(\chi) \equiv M^i_0(\chi)b^i(\chi),
\end{equation}
where we ignore the scale dependence of the bias since we only consider the large-scale signal. Then we get
\begin{equation}\label{E:delta_nuInu}
\delta\left (\nu I_\nu(\nu,\hat{n})  \right )=\sum_{i=1}^{N_c}\int d\chi\, S^i(\nu_{\rm rf})M^i(\chi)A(\chi)\delta_m(\chi, \hat{n}).
\end{equation}

The correlations of $\delta_m$ in the Fourier space is defined by the matter power spectrum $P$,
\begin{equation}\label{E:Pk_matter}
\left \langle \widetilde{\delta}_m(\overrightarrow{k},\chi)\widetilde{\delta}_m^*(\overrightarrow{k'},\chi') \right \rangle 
=\left ( 2\pi \right )^3 \delta^D(\overrightarrow{k}+\overrightarrow{k'})P(k,\chi,\chi').
\end{equation}
In this work, we only consider large scales where the matter power spectrum is scaled by the linear growth rate $G$,
\begin{equation}
P(k,\chi,\chi') = P(k)G(\chi)G(\chi'),
\end{equation}
and we use the linear matter power spectrum at $z=0$ for $P(k)$.

The angular cross power spectrum of frequency $\nu$ and $\nu'$ is
\begin{equation}\label{E:Cl_nunu1}
C_{\ell,\nu\nu'}=\frac{1}{\left (2\ell+1  \right )}\sum_{m=-\ell}^{\ell}\left (a^\nu_{\ell m}  \right )^*a^{\nu'}_{\ell m},
\end{equation}
where $a^{\nu}_{\ell m}$ is the spherical harmonic coefficient of the intensity field
\begin{equation}
\delta\left (\nu I_\nu(\nu,\hat{n})  \right )=\sum_{\ell, m}a^\nu_{\ell m}Y_{\ell m}(\hat{n}).
\end{equation}

We compress the power spectrum into $N_\ell$ bins, where the power spectrum in the $\alpha$th bin, $C_{\ell_\alpha}$, is the averaged $C_\ell$ for modes $\ell\in [\ell_{\alpha,{\rm min}},\ell_{\alpha,{\rm max}})$:
\begin{equation}
C_{\ell_\alpha,\nu\nu'} = \frac{1}{n_{\ell_\alpha}}\sum_{\ell=\ell_{\alpha,{\rm min}}}^{\ell_{\alpha,{\rm max}}-1}\sum_{m=-\ell}^{\ell}\left (a^\nu_{\ell m}  \right )^*a^{\nu'}_{\ell m},
\end{equation}
and the number of modes in the $\alpha$th bin is
\begin{equation}\label{E:number_of_modes}
\begin{split}
n_{\ell_\alpha} &= f_{\rm sky}\sum_{\ell=\ell_{\alpha,{\rm min}}}^{\ell_{\alpha,{\rm max}}-1}\left (2\ell+1  \right )\\
&=f_{\rm sky} \left ( \ell_{\alpha,\rm max}^2-\ell_{\alpha,\rm min}^2 \right ),
\end{split}
\end{equation}
where $f_{\rm sky}$ is the fraction of sky area in the observation. Hereafter, we always consider binning $\ell$ modes in the power spectrum, so we drop the index $\alpha$ for clarity.

We can then write the large-scale (clustering) angular power spectrum as presented in Eq.~\ref{E:Cl_jljl}:
\begin{equation}\label{E:Cl_jljl_appendix}
\begin{split}
C_{\ell,\nu\nu'}^{\rm clus}&= \sum_{i=1}^{N_c}\int d\chi\, S^i(\nu_{\rm rf})M^i(\chi)A(\chi)\\
&\quad\cdot\sum_{{i'}=1}^{N_c}\int d\chi'\, S^{i'}(\nu'_{\rm rf})M^{i'}(\chi')A(\chi')\\
&\quad\cdot\int \frac{dk}{k}\, \frac{2}{\pi}k^3P(k)G(\chi)j_\ell(k\chi)G(\chi')j_\ell(k\chi')
\end{split}
\end{equation}
where $j_\ell$ is the spherical Bessel function. Defining
\begin{equation}\label{E:Bl}
B_\ell^i(k,\nu)=\int d\chi\, S^i(\nu_{\rm rf})M^i(\chi)A(\chi)G(\chi)j_\ell(k\chi)
\end{equation}
and approximating the $k$-integration with Riemann sum in $N_k$ Fourier modes, Eq.~\ref{E:Cl_jljl_appendix} can be rewritten as
\begin{equation}\label{E:Cl_BlBl}
\begin{split}
C_{\ell,\nu\nu'}^{\rm clus}&=\sum_{j=1}^{N_k}\frac{\Delta k_j}{k_j}\frac{2}{\pi}k_j^3P(k_j)\\
&\quad\quad\quad\cdot \left[\sum_{i=1}^{N_c}B_\ell^i(k_j,\nu)\right]\left[ \sum_{i'=1}^{N_c}B_\ell^{i'}(k_j,\nu')\right].
\end{split}
\end{equation}
By arranging the two Bessel functions in Eq.~\ref{E:Cl_jljl} into separate $\chi$-integrations in Eq.~\ref{E:Cl_BlBl}, we can apply the FFTLog algorithm \citep{2017JCAP...11..054A,2018PhRvD..97b3504G,2018JCAP...10..047S} to efficiently evaluate the Bessel function integration. We use the publicly available FFTLog implementation by \citet{2020JCAP...05..010F}\footnote{\url{https://github.com/xfangcosmo/FFTLog-and-beyond}}.

In Eq.~\ref{E:Cl_clus_matrix}, we express the clustering angular power spectrum in the matrix form
\begin{equation}\label{E:Cl_clus_matrix_appendix}
\mathbf{C}^{\rm clus}_\ell = \mathbf{B}_\ell\mathbf{P}\mathbf{B}_\ell^T,
\end{equation}
where $\mathbf{C}^{\rm clus}_\ell$ is an $N_\nu \times N_\nu$ matrix with auto and cross power spectra at mode $\ell$, $\mathbf{B}_\ell$ is an $N_\nu\times N_k$ matrix with elements
\begin{equation}\label{E:Bl_elements}
\mathbf{B}_{\ell,{\nu j}} = \sum_{i=1}^{N_c}B_\ell^i(k_j,\nu),
\end{equation}
and $\mathbf{P}$ is an $N_k\times N_k$ diagonal matrix with elements
\begin{equation}\label{E:P}
\mathbf{P}_{jj} = \frac{\Delta k_j}{k_j}\frac{2}{\pi}k_j^3P(k_j).
\end{equation}

\subsection{Power Spectrum Variance}\label{S:Cl_variance}
The binned power spectrum $\mathbf{C}_\ell$ can be well described by a Gaussian distribution, since each $\ell$ bin contains a large number of independent spherical harmonic coefficients, $a_{\ell m}$, the central limit theorem guarantees its probability distribution converges to a Gaussian in the limit of a large number of samples. Furthermore, as we only consider large scales, the underlying signals are also close to a Gaussian probability density. Therefore, the observed power spectrum $\mathbf{C}_\ell^d$ follows the Wishart distribution with $n_\ell$ (Eq.~\ref{E:number_of_modes}) degree of freedom and scale matrix given by the expected value from model $\mathbf{C}_\ell$. We sample $\mathbf{C}_\ell^d$ independently for each $\ell$ bin, since there is no correlation between multipole modes.

\section{Parametrization}\label{A:parametrization}
\subsection{Basis Functions}\label{S:implementaions_basis}
We define $S^i(\nu_{\rm rf})$ and $M^i(\chi)$ with the linear combination of basis sets $\left\{\hat{S}(\nu_{\rm rf})\right\}$ and $\left\{\hat{M}(\chi)\right\}$ (Eq.~\ref{E:S} and \ref{E:M}). This greatly reduces the computational time on parameter inference. This is because the bottleneck of our algorithm is to evaluate the integration in Eq.~\ref{E:Bl} iteratively during the fitting process. With our parametrization, any $S^i(\nu_{\rm rf})$ and $M^i(\chi)$ can be written as their linear combination with coefficient sets $\{c^i_{S,m}\}$ and $\{c^i_{M,n}\}$, respectively. Therefore, we can precompute Eq.~\ref{E:Bl} integration for all combinations of basis at each frequency band $\nu$ and Fourier bin $k_j$,
\begin{equation}
\hat{B}_{\ell,mn}(k_j,\nu)=\int d\chi\, A(\chi)\hat{M}_n(\chi)\hat{S}_m(\nu_{\rm rf})G(\chi)j_\ell(k_j\chi),
\end{equation}
and then obtain $B^i_\ell$ (Eq.~\ref{E:Bl}) for a given $S^i$ and $M^i$ with the linear combination
\begin{equation}\label{E:Bl_basis}
B^i_\ell(k_j,\nu) = \sum_{m=1}^{N_s}\sum_{n=1}^{N_m} c^i_{S,m}c^i_{M,n}\hat{B}_{\ell,mn}(k_j,\nu).
\end{equation}
With this setup, the integration in Eq.~\ref{E:Bl} only needs to be evaluated $N_\ell\times N_\nu\times N_k \times N_s \times N_m$ times for all combinations of basis before fitting to the observed power spectra $\mathbf{C}^d_\ell$.

\section{Poisson Noise}\label{A:Poisson}
The Poisson noise of the cross angular power spectrum $C_{\ell,\nu\nu'}$ from sources is 
\begin{equation}
\begin{split}
C_{\ell,\nu\nu'}^{\rm P}&=\sum_{i=1}^{N_c}\int d\chi\int dL^i\,\Phi^i(L^i,\chi)D_A^2(\chi)\\
&\quad\quad\cdot\left[
\frac{\nu_{\rm rf}L^i_\nu(\nu_{\rm rf})}{4\pi D_L^2(\chi)}\right]\left[
\frac{\nu'_{\rm rf}L^i_\nu(\nu'_{\rm rf})}{4\pi D_L^2(\chi)}\right]\\
&=\sum_{i=1}^{N_c}\int d\chi\int dL^i\,(L^i)^2\Phi^i(L^i,\chi)D_A^2(\chi)\\
&\quad\quad\cdot\left[
\frac{\nu_{\rm rf}L^i_\nu(\nu_{\rm rf})/L^i}{4\pi D_L^2(\chi)}\right]\left[
\frac{\nu'_{\rm rf}L^i_\nu(\nu'_{\rm rf})/L^i}{4\pi D_L^2(\chi)}\right],
\end{split}
\end{equation}
where $\nu_{\rm rf}=(1+z)\nu$, and $\nu'_{\rm rf}=(1+z)\nu'$. Defining the Poisson-to-clustering ratio,
\begin{equation}
r^i_{{\rm P}}(\chi) \equiv \frac{\int dL\,L^2\Phi^i(L,\chi)}{\left[b^i(\chi)\int dL\,L\Phi^i(L,\chi)\right]^2} = \frac{\int dL\,L^2\Phi^i(L,\chi)}{\left[M^i(\chi)\right]^2},
\end{equation}
and a redshift-dependent factor similar to $A(\chi)$ in the clustering case (Eq.~\ref{E:SMP_def}),
\begin{equation}
A_P(\chi)\equiv\frac{D_A(\chi)}{4\pi D_L^2(\chi)},
\end{equation}
we can express the cross Poisson noise as
\begin{equation}
C_{\ell,\nu\nu'}^{\rm P} = \sum_{i=1}^{N_c}\int d\chi\, r^i_{\rm P}(\chi)S^i(\nu_{\rm rf})S^i(\nu'_{\rm rf})\left[M^i(\chi)\right]^2 A_{\rm P}^2(\chi).
\end{equation}
Therefore, with $S^i$ and $M^i$ from the clustering power spectrum, we can model the Poisson noise by characterizing the $r^i_{\rm P}(\chi)$ function for each component $i$.

\section{Regularization}\label{A:regularization}
To break the amplitude degeneracy of $S^i$, $M^i$, and $P$, we define the following regularization term in the prior to fix the overall scaling of $S^i$ and $P$,
\begin{equation}\label{E:prior_reg}
\begin{split}
{\rm log}\,\pi_{\rm reg}(\mathbf{\Theta})=&- \lambda_S\, \,\frac{1}{N_c}\sum_{i=1}^{N_c}\left [ \left ( \sum_{m=1}^{N_s} c^i_{S,m} \right )-1 \right ]^2 \\
&- \lambda_P \,\frac{1}{N_k}\sum_{j=1}^{N_k}\left ( \frac{P(k_{j})}{P^{\rm fid}(k_{j})} -1\right )^2,
\end{split}
\end{equation}
where $P^{\rm fid}$ is the fiducial model of the matter power spectrum, and $\lambda_S$ and $\lambda_P$ are the regularization strengths, and we use 
\begin{equation}
\lambda_S=\lambda_P=0.1\cdot \left[-\frac{1}{2}\sum_\ell n_\ell\, {\rm log}\,\mathcal{N}\left ( \mathbf{C}^d_\ell,\mathbf{C}^d_\ell \right )\right],
\end{equation}
where $\mathcal{N}$ is the normal distribution (see Eq.~\ref{E:likelihood}) We check that with our choice of regularization strength ($\lambda_S$ and $\lambda_P$), $p_{\rm reg}$ is relatively flat compared to the likelihood $\mathcal{L}$ at the fiducial parameter values, and thus this additional regularization term will not bias the posterior inference.

From Eq.~\ref{E:prior_reg} we can derive the regularization term in the Fisher matrix (Eq.~\ref{E:Fisher_inv}),
\begin{equation}
\begin{split}
&\mathbf{F}_{{\rm reg},\alpha\beta} = -\left \langle \frac{\partial^2 {\rm log}\,p_{\rm reg}}{\partial\theta_\alpha\partial\theta_\beta} \right \rangle\\
&=
\begin{cases}
\frac{2\lambda_S}{N_c} \,\, \text{if}\,\,
\theta_\alpha, \theta_\beta=c^i_{S,m}, c^i_{S,m'}\\
\frac{2\lambda_P}{N_k P^{\rm fid}(k_j)P^{\rm fid}(k_{j'})} \,\, \text{if}\,\,
\theta_\alpha, \theta_\beta=P(k_j), P(k_{j'})\\
0 \quad\quad \text{otherwise}
\end{cases}
\end{split}
\end{equation}

\section{Newton-Raphson Method}\label{A:Newtons}
The Newton-Raphson method is an iterative method to find the minimum/maximum of a function. Here, we seek for the solution $\Theta_{\rm max}$ that gives the maximum log-likelihood ${\rm ln}\,\mathcal{L}$. Using the Newton-Raphson algorithm, at step $t+1$, we update the parameter set from $\Theta_{t}$ to $\Theta_{t+1}$ by
\begin{equation}
\Theta_{t+1}=\Theta_{t} - \eta \mathbf{H}^{-1}\mathbf{g},
\end{equation}
where $\eta$ is the learning rate, the gradient $\mathbf{g}$ is an $N_\theta$-sized vector with elements
\begin{equation}
\begin{split}
\mathbf{g}_\alpha&=\frac{\partial{\rm ln}\,\mathcal{L}}{\partial \theta_\alpha}\\
&=-\frac{1}{2}\sum_\ell n_\ell{\rm Tr}\left [ \left ( -\mathbf{C}_\ell^{-1}\mathbf{C}^d_\ell\mathbf{C}_\ell^{-1} + \mathbf{C}_\ell^{-1} \right )\frac{\partial \mathbf{C}_\ell}{\partial\theta_\alpha}\right ],
\end{split}
\end{equation}
and the Hessian $\mathbf{H}$ is an $N_\theta\times N_\theta$ matrix with elements
\begin{equation}
\begin{split}
&\mathbf{H}_{\alpha\beta}=\frac{\partial^2 {\rm ln}\,\mathcal{L}}{\partial\theta_\alpha\partial\theta_\beta}\\
=&-\frac{1}{2}\sum_\ell n_\ell \frac{\partial^2 }{\partial\theta_\alpha\partial\theta_\beta}\left [\right. {\rm Tr}\left ( \mathbf{C}^d_\ell \mathbf{C}_\ell^{-1}\right )+{\rm log\,det}\left ( \mathbf{C}_\ell\right )\left.\right ]\\
=&-\frac{1}{2}\sum_\ell n_\ell\frac{\partial}{\partial\theta_\alpha}{\rm Tr}\left [ \left ( -\mathbf{C}_\ell^{-1}\mathbf{C}^d_\ell\mathbf{C}_\ell^{-1} + \mathbf{C}_\ell^{-1} \right )\frac{\partial \mathbf{C}_\ell}{\partial\theta_\beta}\right ]\\
=&-\frac{1}{2}\sum_\ell n_\ell{\rm Tr}\left [ \frac{\partial}{\partial\theta_\alpha}\left ( -\mathbf{C}_\ell^{-1}\mathbf{C}^d_\ell\mathbf{C}_\ell^{-1} + \mathbf{C}_\ell^{-1} \right )\frac{\partial \mathbf{C}_\ell}{\partial\theta_\beta}\right ]\\
&-\frac{1}{2}\sum_\ell n_\ell{\rm Tr}\left [ \left ( -\mathbf{C}_\ell^{-1}\mathbf{C}^d_\ell\mathbf{C}_\ell^{-1} + \mathbf{C}_\ell^{-1} \right )\frac{\partial^2 \mathbf{C}_\ell}{\partial\theta_\alpha\partial\theta_\beta}\right ].
\end{split}
\end{equation}
Using
\begin{equation}
\begin{split}
&\frac{\partial}{\partial\theta_\alpha}\left ( -\mathbf{C}_\ell^{-1}\mathbf{C}^d_\ell\mathbf{C}_\ell^{-1} + \mathbf{C}_\ell^{-1} \right ) \\
&= -\frac{\partial\mathbf{C}_\ell^{-1}}{\partial\theta_\alpha}\mathbf{C}^d_\ell\mathbf{C}_\ell^{-1}-\mathbf{C}_\ell^{-1}\mathbf{C}^d_\ell\frac{\partial\mathbf{C}_\ell^{-1}}{\partial\theta_\alpha}-\frac{\partial\mathbf{C}_\ell^{-1}}{\partial\theta_\alpha}\\
&=\mathbf{C}_\ell^{-1}\frac{\partial\mathbf{C}_\ell}{\partial\theta_\alpha}\mathbf{C}_\ell^{-1}\mathbf{C}^d_\ell\mathbf{C}_\ell^{-1}+\mathbf{C}_\ell^{-1}\mathbf{C}^d_\ell\mathbf{C}_\ell^{-1}\frac{\partial\mathbf{C}_\ell}{\partial\theta_\alpha}\mathbf{C}_\ell^{-1}\\
&\quad\quad\quad\quad\quad\quad\quad\quad\quad\quad-\mathbf{C}_\ell^{-1}\frac{\partial\mathbf{C}_\ell}{\partial\theta_\alpha}\mathbf{C}_\ell^{-1}\\
&=2\mathbf{C}_\ell^{-1}\frac{\partial\mathbf{C}_\ell}{\partial\theta_\alpha}\mathbf{C}_\ell^{-1}\mathbf{C}^d_\ell\mathbf{C}_\ell^{-1}-\mathbf{C}_\ell^{-1}\frac{\partial\mathbf{C}_\ell}{\partial\theta_\alpha}\mathbf{C}_\ell^{-1},
\end{split}
\end{equation}
we get
\begin{equation}\label{E:H_exact}
\begin{split}
\mathbf{H}_{\alpha\beta}=&-\frac{1}{2}\sum_\ell n_\ell\left[ {\rm Tr}\left ( 2\mathbf{C}_\ell^{-1}\frac{\partial\mathbf{C}_\ell}{\partial\theta_\alpha}\mathbf{C}_\ell^{-1}\mathbf{C}^d_\ell\mathbf{C}_\ell^{-1}\frac{\partial \mathbf{C}_\ell}{\partial\theta_\beta}   \right.\right.\\
&\left.\left.\quad\quad\quad\quad\quad\quad\quad\quad-\mathbf{C}_\ell^{-1}\frac{\partial \mathbf{C}_\ell}{\partial\theta_\alpha}\mathbf{C}_\ell^{-1}\frac{\partial \mathbf{C}_\ell}{\partial\theta_\beta} \right )\right] \\
&-\frac{1}{2}\sum_\ell n_\ell\left[{\rm Tr} \left ( -\mathbf{C}_\ell^{-1}\mathbf{C}^d_\ell\mathbf{C}_\ell^{-1}\frac{\partial^2 \mathbf{C}_\ell}{\partial\theta_\alpha\partial\theta_\beta}\right.\right.\\
&\quad\quad\quad\quad\quad\quad\quad\quad+\left.\left.\mathbf{C}_\ell^{-1}\frac{\partial^2 \mathbf{C}_\ell}{\partial\theta_\alpha\partial\theta_\beta} \right )\right].
\end{split}
\end{equation}
The $\mathbf{C}_\ell$'s derivatives on parameters $\frac{\partial\mathbf{C}_\ell}{\partial\theta}$ are given in Appendix~\ref{A:dCl_dtheta}.

When implementing the Newton-Raphson method, instead of using the exact expression of Eq.~\ref{E:H_exact}, we use the approximated Hessian
\begin{equation}
\hat{\mathbf{H}}_{\alpha\beta} = -\frac{1}{2}\sum_\ell n_\ell {\rm Tr}\left ( \mathbf{C}_\ell^{-1}\frac{\partial\mathbf{C}_\ell}{\partial\theta_\alpha}\mathbf{C}_\ell^{-1}\frac{\partial\mathbf{C}_\ell}{\partial\theta_\beta} \right ),
\end{equation}
which approaches the exact expression (Eq.~\ref{E:H_exact}) when $\mathbf{C}_\ell^d\rightarrow \mathbf{C}_\ell$. The approximation helps us to avoid evaluating $O(N_\ell N_\theta^2)$ number of second derivatives on all parameters $\frac{\partial^2 \mathbf{C}_\ell}{\partial\theta_\alpha\partial\theta_\beta}$, and therefore we can greatly speed up the Newton-Raphson iterations.

In each step, we adjust the learning rate to guarantee an increment of ${\rm ln}\,\mathcal{L}$ after updating the parameters. 

Further implementation details on applying the Newton-Raphson method to our problem will be presented in future papers.

\section{Fisher Matrix}\label{A:Fisher}
The Fisher matrix is the expectation value of the inverse Hessian,
\begin{equation}
\mathbf{F}_{\alpha\beta} = -\left \langle \frac{\partial^2 {\rm log}\,\mathcal{L}}{\partial\theta_\alpha\partial\theta_\beta} \right \rangle.
\end{equation}
Since $\left< \mathbf{C}^d_\ell\right>= \mathbf{C}_\ell$, the second term in Eq.~\ref{E:H_exact} vanishes, and therefore,
\begin{equation}
\mathbf{F}_{\alpha\beta} = \frac{1}{2}\sum_\ell n_\ell {\rm Tr}\left ( \mathbf{C}_\ell^{-1}\frac{\partial\mathbf{C_\ell}}{\partial\theta_\alpha}\mathbf{C}_\ell^{-1}\frac{\partial\mathbf{C_\ell}}{\partial\theta_\beta} \right ).
\end{equation}

\def\Cl{$\mathbf{C}_\ell$}
\section{\Cl ~Derivatives}\label{A:dCl_dtheta}
Both the Newton-Raphson method and the Fisher matrix calculations require the derivatives of $\mathbf{C}_\ell$ on parameters $\mathbf{\Theta}=\left \{ \left \{ c^i_{S,m} \right \}, \left \{ c^i_{M,n} \right \}, \left \{ P(k_j) \right \}, \left \{ \mathbf{N}_{\ell_\alpha,\nu\nu}\right \} \right \}$. With our parametrization, we can analytically express $\partial\mathbf{C_\ell}/\partial\theta$ for all parameters. From Eq.~\ref{E:Cl} and using the fact that $\left \{ c^i_{S,m} \right \}$ and $ \left \{ c^i_{M,n} \right \}$ only depend on $\mathbf{B}_\ell$, $\left \{ P(k_j) \right \}$ only depends on $\mathbf{P}$, and $ \left \{ \mathbf{N}_{\ell_\alpha,\nu\nu}\right \}$ only depends on $\mathbf{N}_\ell$, we can write the power spectrum derivatives as
\begin{align}
\frac{\partial\mathbf{C_\ell}}{\partial c^i_{S,m}}&=\frac{\partial\mathbf{B_\ell}}{\partial c^i_{S,m}}\mathbf{P}\mathbf{B}_\ell^T + \mathbf{B}_\ell\mathbf{P}\frac{\partial\mathbf{B_\ell}}{\partial c^i_{S,m}}^T,\\
\frac{\partial\mathbf{C_\ell}}{\partial c^i_{M,n}}&=\frac{\partial\mathbf{B_\ell}}{\partial c^i_{M,n}}\mathbf{P}\mathbf{B}_\ell^T + \mathbf{B}_\ell\mathbf{P}\frac{\partial\mathbf{B_\ell}}{\partial c^i_{M,n}}^T,\\
\frac{\partial\mathbf{C_\ell}}{\partial P(k_j)}&=\mathbf{B}_\ell\frac{\partial\mathbf{P}}{\partial P(k_j)}\mathbf{B}_\ell^T,\\
\frac{\partial\mathbf{C_\ell}}{\partial \mathbf{N}_{\ell_\alpha,\nu\nu}}&=
\frac{\partial\mathbf{N_\ell}}{\partial \mathbf{N}_{\ell_\alpha,\nu\nu}}.
\end{align}
With our basis function expansion (Eq.~\ref{E:Bl_elements} and \ref{E:Bl_basis}), we get
\begin{equation}
\begin{split}
&\frac{\partial\mathbf{B}_{\ell,\nu j}}{\partial c^i_{S,m}}
=\frac{\partial}{\partial c^i_{S,m}}\sum_{i'=1}^{N_c}B_\ell^{i'}(k_j,\nu)=\frac{\partial B_\ell^i(k_j,\nu)}{\partial c^i_{S,m}}\\
&=\frac{\partial}{\partial c^i_{S,m}}\left[\sum_{m'=1}^{N_s}\sum_{n'=1}^{N_m} c^i_{S,m'}c^i_{M,n'}\hat{B}_{\ell,m'n'}(k_j,\nu)\right]\\
&= \sum_{n=1}^{N_m} c^i_{M,n}\hat{B}_{\ell,mn}(k_j,\nu),
\end{split}
\end{equation}
and similarly,
\begin{equation}
\frac{\partial\mathbf{B}_{\ell,\nu j}}{\partial c^i_{M,n}}
= \sum_{m=1}^{N_s} c^i_{S,m}\hat{B}_{\ell,mn}(k_j,\nu).
\end{equation}
The derivatives of $\mathbf{P}$ and $\mathbf{N}_\ell$ are 

\begin{equation}
\frac{\partial\mathbf{P}}{\partial P(k_j)}=\delta^K_{jj},
\end{equation}
\begin{equation}
\frac{\partial\mathbf{N_\ell}}{\partial \mathbf{N}_{\ell_\alpha,\nu\nu}}=\delta^K_{\ell\ell_\alpha}\delta^K_{\nu\nu},
\end{equation}
where $\delta^K$ is the Kronecker delta.

\section{MCMC Implementation}\label{A:MCMC}
We use MCMC to verify the results from the Newton-Raphson method and the Fisher matrix. To help the MCMC sampler converge more efficiently, we add another two terms to the prior:
\begin{equation}
\pi\left ( \mathbf{\Theta}\right ) = \pi_{\rm J}\left ( \mathbf{\Theta}\right ) \pi_{
\rm lim}\left ( \mathbf{\Theta}\right )\pi_{\rm reg}\left ( \mathbf{\Theta}\right ).
\end{equation}
The first term, $\pi_{\rm J} $, is a Jeffreys prior ($\pi(\theta)\propto 1/\theta$) on the $P$ and $\mathbf{N}_{\ell}$ parameters to better sample the potentially unknown scales of these parameters:
\begin{equation}
\pi_{\rm J}\left ( \mathbf{\Theta}\right )=\left [ \prod _{j=1}^{N_k}\frac{1}{P(k_j)} \right ]\cdot\left [ \prod _{\ell=1}^{N_\ell}\prod _{\nu=1}^{N_\nu}\frac{1}{\mathbf{N}_{\ell,\nu\nu}} \right ].
\end{equation}
We use flat priors for $\left \{ c^i_{S,m} \right \}$ and $\left \{ c^i_{M,n} \right \}$. The second term, $\pi_{\rm lim}\left ( \mathbf{\Theta}\right )$, is used to impose limits on the parameters. Here, we require all $S^i$ coefficients ($c^i_{S,m}$) and the $M^i(\chi)$ function to be non-negative\footnote{We found the MCMC fitting converges better by setting stronger positivity constraints, $c^i_{S,m} \geq 0$, instead of $S^i = \sum_{m=1}^{N_s} c^i_{S,m}\,\hat{S}_m(\nu_{\rm rf}) \geq 0$.}, and the $P(k)$ and $\mathbf{N}_\ell$ are confined to a range. Therefore, we set $\pi_{\rm lim}\left ( \mathbf{\Theta}\right )=1$, if
\begin{equation}\label{S:prior_lim}
\begin{cases}
&c^i_{S,m} \geq 0,\\
&M^i= \sum_{n=1}^{N_m} c^i_{M,n}\,\hat{M}_n(\chi)\geq0,\\
&P^{\rm min}(k_j) < P(k_j) < P^{\rm max}(k_j),\\
&\mathbf{N}^{\rm min}_{\ell,\nu\nu} < \mathbf{N}_{\ell_\alpha,\nu\nu} < \mathbf{N}^{\rm max}_{\ell,\nu\nu},
\end{cases}
\end{equation}
and $\pi_{\rm lim}\left ( \mathbf{\Theta}\right )=0$, otherwise. We set $P$ and $\mathbf{N}_{\ell}$ to $\pm 50\%$ and $\pm 10\%$ of the fiducial input values $P^{\rm fid}$ and $\mathbf{N}_{\ell}^{\rm fid}$, respectively. 

With a large number of parameters ($N_\theta$), common Metropolis--Hasting algorithm implementations are inefficient, due to the low acceptance rate. Therefore, we use the blocked Gibbs sampling method, which only updates a subset of parameters at a time to get faster convergence. We divide parameters into $N_\ell+1$ blocks: $\mathbf{\Theta} = \left \{ \mathbf{\Theta}_{\rm SMP}, \mathbf{\Theta}_{N_{\ell_1}}, \mathbf{\Theta}_{N_{\ell_2}},... \right \}$, where
$\mathbf{\Theta}_{SMP} = \left \{ \left \{ c^i_{S,m} \right \}, \left \{ c^i_{M,n} \right \}, \left \{ P(k_j) \right \} \right \}$ and $\mathbf{\Theta}_{N_{\ell_\alpha}} = \left \{ \mathbf{N}_{\ell_\alpha,\nu\nu}\right \}$. At step $t$, the blocked Gibbs sampler draws a new sample $\mathbf{\Theta}^{t+1}$ from the current parameter values $\mathbf{\Theta}^{t}$ one block at a time by sampling from the conditional distribution. We first sample $\mathbf{\Theta}_{\rm SMP}$ by
\begin{equation}\label{E:theta_SMP}
\mathbf{\Theta}_{\rm SMP}^{t+1} \leftarrow p\left ( \mathbf{\Theta}_{\rm SMP}|\mathbf{\Theta}_{N_{\ell_1}}^t,\mathbf{\Theta}_{N_{\ell_2}}^t,...,\left\{\mathbf{C}^d_\ell\right\}\right ),
\end{equation}
and then update $\mathbf{\Theta}_{N_{\ell_{1}}}^{t+1}$, $\mathbf{\Theta}_{N_{\ell_{2}}}^{t+1}$, ... by
\begin{equation}\label{E:theta_N}
\begin{split}
\mathbf{\Theta}_{N_{\ell_{\alpha}}}^{t+1} \leftarrow p\left (\right. &\mathbf{\Theta}_{N_{\ell_{\alpha}}}|\mathbf{\Theta}_{SMP}^{t+1}, \mathbf{\Theta}_{N_{\ell_1}}^{t+1},\mathbf{\Theta}_{N_{\ell_2}}^{t+1},...,\\
&\mathbf{\Theta}_{N_{\ell_{\alpha-1}}}^{t+1},\mathbf{\Theta}_{N_{\ell_{\alpha+1}}}^t,...,\left\{\mathbf{C}^d_\ell\right\}  \left.\right )
\end{split}
\end{equation}
We note that since $\ell$ modes are independent in the likelihood, we can sample $\mathbf{\Theta}_{N_{\ell_{\alpha}}}^{t+1}$ for each $\ell$ mode simultaneously from the conditional distribution $p\left ( \right.\mathbf{\Theta}_{N_{\ell_{\alpha}}}|\mathbf{\Theta}_{SMP}^{t+1},\mathbf{C}^d_\ell \left. \right )$. We use the affine-invariant MCMC sampler \texttt{emcee} \citep{2013PASP..125..306F} to sample from the conditional probability distribution (Eq.~\ref{E:theta_SMP} and Eq.~\ref{E:theta_N}).

\bibliography{cosmo3D}{}
\bibliographystyle{aasjournal}

\end{document}